\documentclass[12pt]{iopart}

\expandafter\let\csname equation*\endcsname\relax
  \expandafter\let\csname endequation*\endcsname\relax

\usepackage{graphicx}

\usepackage[utf8]{inputenc}
\usepackage[T1]{fontenc}
\usepackage{mathtools, bm}
\usepackage{amsmath}
\usepackage{amsmath,amssymb}  % Better maths support & more symbols
\usepackage{textcomp} % provide lots of new symbols
\usepackage[font=small]{caption}
\usepackage{mathrsfs}
\DeclareSymbolFont{calletters}{OMS}{cmsy}{m}{n}
\DeclareSymbolFontAlphabet{\mathcal}{calletters}
\usepackage{calligra}
\DeclareMathAlphabet{\mathcalligra}{T1}{calligra}{m}{n}

\usepackage{comment}
\usepackage{color}

\newcommand{\para}{\mathbin{\|}}
\newcommand{\pa}{{\para}}
\newcommand{\pe}{{\perp}}

\newcommand{\F}{\mathcal{F}}

\newcommand{\G}{\mathcal{G}}

\newcommand{\bstar}{b_{\star}}

\newcommand{\apar}{A_\parallel}
\newcommand{\bpar}{B_\parallel}
\newcommand{\lapp}{\Delta_\perp}
\newcommand{\bx}{\mathbf{x}}
\newcommand{\scrt}{\mathscr{T}}
\newcommand{\scrd}{\mathscr{D}}
\newcommand{\bk}{\mathbf{k}}
\newcommand{\beq}{\begin{equation}}
\newcommand{\eeq}{\end{equation}}
\newcommand{\nno}{\nonumber}
\newcommand{\dn}{\mathcal{D}_n}
\newcommand{\wbpar}{\widetilde{B}_\parallel}
\newcommand{\wapar}{ \widetilde{A}_{\parallel}}
\newcommand{\wphi}{\widetilde{\phi}}
\newcommand{\Tpa}{T_{{0 }_{\parallel s}}}

\newcommand{\tcalfa}{\widetilde{\mathcal{F}}_{{0s}}}
\newcommand{\ga}{\widetilde{g}_s}

\newcommand{\aal}{a_{s}}

\newcommand{\qa}{q_{s}}
\newcommand{\dwa}{d \mathcal{W}_s}
\newcommand{\Tpea}{T_{{0 }_{\perp s}}}
\newcommand{\thea}{\Theta_s}
\newcommand{\vtpa}{v_{{th }_{\parallel s}}}
\newcommand{\bepea}{\beta_{\perp_s}}
\newcommand{\dfa}{\widetilde{f}_{s}}
\newcommand{\tpi}{\tau_{\perp_i}}
\newcommand{\ee}{\mathrm{e}}
\newcommand{\rthi}{\rho_{th_{\pe i}}}
\newcommand{\rhos}{\rho_{s \perp}}
\newcommand{\bpe}{\beta_{\perp_e}}
\newcommand{\bd}{{\partial D_n}}
\newcommand{\bn}{\mathbf{n}}
\newcommand{\hz}{\hat{z}}
\newcommand{\intd}{\int_{D_n}}
\newcommand{\intbd}{\int_\bd}
\newcommand{\dl}{d \mathbf{l}}
\newcommand{\km}{k_{min}^2}
\newcommand{\kr}{k_R^2}
\newcommand{\ueb}{\mathbf{U}_{\mathbf{E}\times\mathbf{B}}}

\newcommand{\tcalfe}{\widetilde{\mathcal{F}}_{{0e}}}

\newcommand{\dfe}{\widetilde{f}_e}
\newcommand{\vtpe}{v_{th \parallel e}}
\newcommand{\Tpe}{T_{0 \parallel e}}
\newcommand{\Tpee}{T_{0 \perp e}}
\newcommand{\tpe}{T_{\parallel e}}
\newcommand{\tpee}{T_{\perp e}}
\newcommand{\dwe}{d \mathcal{W}_e}
\newcommand{\dwi}{d \mathcal{W}_i}
\newcommand{\dfi}{\widetilde{f}_i}
\newcommand{\lbphi}{\bar{\mathcal{L}}_\phi}
\newcommand{\lbb}{\bar{\mathcal{L}}_B}
\newcommand{\lba}{\bar{\mathcal{L}}_A}
\newcommand{\lphi}{\mathcal{L}_\phi}
\newcommand{\lb}{\mathcal{L}_B}

\newcommand{\lpar}{L_\parallel}

\newcommand{\tx}{\tilde{x}}
\newcommand{\ty}{\tilde{y}}
\newcommand{\tz}{\tilde{z}}
\newcommand{\tv}{\tilde{v}_\parallel}
\newcommand{\tmus}{\tilde{\mu}_{0s}}
\newcommand{\tmue}{\tilde{\mu}_{0e}}
\newcommand{\tit}{\tilde{t}}

\begin{document}

\title[Linear stability of magnetic vortex chains in a plasma]{Linear stability of magnetic vortex chains in a plasma in the presence of equilibrium electron temperature anisotropy}

\author{C. Granier$^{1}$, E. Tassi$^{1}$}
\address{$^1$ Universit\'e C\^ote d'Azur, CNRS, Observatoire de la C\^ote d'Azur, Laboratoire J. L. Lagrange, Boulevard de l'Observatoire, CS 34229, 06304 Nice Cedex 4, France}

\baselineskip 24 pt

\maketitle

\begin{abstract}
The linear stability of chains of magnetic vortices in a plasma is investigated analytically in two dimensions by means of a reduced fluid model assuming a strong guide field and accounting for equilibrium electron temperature anisotropy. The chain of magnetic vortices is modelled by means of the classical "cat's eyes" solutions and the linear stability is studied by analysing the second variation of a conserved functional, according to the Energy-Casimir method. 
The stability analysis  is carried out on the domain bounded by the separatrices of the vortices.  Two cases are considered, corresponding to a ratio between perpendicular equilibrium ion and electron temperature much greater or much less than unity, respectively. In the former case, equilibrium flows depend on an arbitrary function. Stability is attained if the equilibrium electron temperature anisotropy is bounded from above and from below, with the lower bound corresponding to the condition preventing the firehose instability. A further condition sets an upper limit to the amplitude of the vortices, for a given choice of the equilibrium flow. For cold ions, two sub-cases have to be considered. In the first one, equilibria correspond to those for which the velocity field is proportional to the local Alfv\'en velocity. Stability conditions imply: an upper limit on the amplitude of the flow, which automatically implies firehose stability, an upper bound on the electron temperature anisotropy and again an upper bound on the size of the vortices. The second sub-case refers to equilibrium electrostatic potentials which are not constant on magnetic flux surfaces and the resulting stability conditions correspond to those of the first sub-case in the absence of flow.

\end{abstract}

\section{Introduction}  \label{sec:intro}

The identification of coherent structures and the investigation of their stability is a classical subject in plasma physics. Among the various coherent structures that can form in plasmas, chains of magnetic vortices  (also referred to as magnetic islands) are of considerable relevance for both laboratory and space plasmas, and the study of their stability began already a few decades ago \cite{Fin77,Pri79,Bis80,Bon83,Bon83b}. Such stability analysis  was often carried out in the context of a magnetohydrodynamic (MHD) description of a plasma and the modelling of the magnetic vortex chain took advantage from the existence of a well known solution of the Liouville equation (explicitly given later in Eq. (\ref{ceyes}))
which can be applied when investigating plasma equilibria with a symmetry \cite{Fad65}. This solution was adopted much earlier in fluid dynamics, where it is usually referred to as Kelvin-Stuart "cat's eyes" solution \cite{Kel80,Stu67}.  In plasma physics, such equilibrium solution proved to be a standard starting point for the investigation of problems related to island coalescence (see, for instance Refs. \cite{Sta15,Pri00,Bis00,Taj91} and references therein). To the best of our knowledge, however, analytical investigations of the stability of magnetic island chains remained a minority, with respect to the vast amount of numerical results obtained on this subject.  In particular, the impact of some two-fluid effects on the stability of magnetic vortex chains seems to lack a fully analytical description. 

The purpose of this paper is to provide, by means of fully analytical methods, sufficient conditions for the linear stability of classes of equilibria with "cat's eyes" solutions for the magnetic field, in the framework of a reduced fluid model accounting for two-fluid effects. More precisely, we consider equilibrium solutions of the model, for which the magnetic field, in the plane perpendicular to a constant and uniform guide field, is described by the "cat's eyes" solution. The adopted reduced fluid model can be derived from the set of gyrokinetic equations described in Ref. \cite{kunz2015}. Its Hamiltonian structure can be derived from that of the two-field Hamiltonian gyrofluid model considered in Ref. \cite{Tas19} (see in particular Sec. 6 of such Reference), considering the two-dimensional (2D) limit and letting go to zero the electron-to-ion mass ratio.  In its general formulation, the model accounts also for ion finite Larmor radius (FLR) effects. For this reason, in Sec. \ref{sec:model}, we refer to it as to a gyrofluid model. In the present analysis, however, only two extreme and opposite limits will be considered, i.e. when the ion temperature, referred to the plane perpendicular to the guide field, is much greater and much less than the corresponding electron temperature, respectively. 

The investigation of the above mentioned two-fluid effects could shed some light, for instance, on instabilities driven by electron temperature anisotropy on magnetic vortex chains. This mechanism might be relevant for nearly collisionless plasmas, such as the solar wind, where the equilibrium distribution functions of particle populations are typically anisotropic.  With regard to this, we remark a recent application of the "cat's eyes" solutions, in the framework of reduced MHD \cite{Jov18}, for the description of magnetic vortex chains observed in the solar wind by the Cluster spacecraft \cite{perrone2016,perrone2017}. Observational data proved indeed to yield structures  compatible with those of the "cat's eyes" solution. Such analysis, however, focused on scales larger than the ion thermal gyroradius, where two-fluid effects have little relevance. 

With regard to ion temperature effects, although our analysis is limited to two extreme cases, it might provide a leading order indication of what configurations of electron gyrocenter density, electrostatic potentials and parallel magnetic perturbations can support magnetic vortex chains in plasmas with hot or cold ions at equilibrium (or, equivalently, at scales smaller or larger than the ion thermal gyroradius, given that the characteristic scale of the model is the perpendicular sonic Larmor radius).

The method adopted for the stability analysis is the Energy-Casimir method for determining formal stability, which implies linear stability \cite{Mor98,Hol85}. This method typically applies to Hamiltonian systems with a noncanonical Poisson bracket and is based on identifying conditions for which the second variation of a functional conserved by the model has a definite sign, when evaluated at the equilibrium point. This method is described in Refs. \cite{Mor98,Hol85} and examples of its application in the fluid and plasma physics literature can be found in Refs. \cite{Fjo50,Kru58,Mor90,Mor13,And13,And16}. An application of this method to a plasma equilibrium with a "cat's eyes" chain of vortices is provided in Ref. \cite{dagnelund}. We also point out the description, in Ref. \cite{Thr09}, of an MHD analytical investigation of the stability of magnetic vortex chains in the presence of flows, with application to tokamaks. 

We mention that the steps of the Energy-Casimir method for linear stability analysis adopted here, can be extended to yield conditions for nonlinear stability, by carrying out further estimates. This procedure is applied to fluids in Refs. \cite{Arn65,Arn69} and is described with various fluid and plasma examples in Ref. \cite{Hol85}. In Ref. \cite{Hol86b}, an analysis based on this method yields conditions for nonlinear stability of the "cat's eyes" solution for the 2D Euler equation for an incompressible fluid. In the absence of results on the existence of solutions for the nonlinear system under investigation, which is the case for the present model, however, only conditional nonlinear stability can be proved \cite{Hol85}. Therefore, we content with deriving conditions for linear stability, which is also what is provided by usual analytical stability methods adopted in plasma physics.

Finally, we remark that a technical difficulty posed by the present problem concerns the two-dimensional (2D) domain where the stability analysis is carried out. We choose this domain to be the portion of space enclosed by the separatrices of the vortices, borrowing a procedure adopted in Ref. \cite{Hol86b}.

The paper is organized as follows. In Sec. \ref{sec:model} we introduce the reduced gyrofluid model valid for arbitrary ion temperature, review its main properties and present its Hamiltonian structure. The Casimir invariants of the model (which are analogous to those of 2D reduced MHD) are recalled and, together with the Hamiltonian, will form the starting point for the stability analysis. At the end of the Section we introduce the spatial domain where the stability analysis will be carried out. Sections \ref{sec:hions} and \ref{sec:cions} present the stability analysis in the limit of hot and cold ions, respectively. Both Sections begin with the introduction of the model equations in the corresponding limit, and of their conserved quantities. This is followed by the analysis of the first variation of the conserved functional, which leads to the classes of equilibria of interest. The two Sections end with the analysis of the second variation, yielding the stability conditions, which are discussed in the final part of each Section.  We conclude in Sec. \ref{sec:concl}. Two Appendices are also provided. In  \ref{sec:gyrokin} the physical assumptions and the main steps in the derivation of the model, starting from the gyrokinetic equations, are presented. The relation between the Hamiltonian structure of the present model and that of the more general three-dimensional (3D) gyrofluid model presented in Ref. \cite{Tas19} is also discussed. \ref{sec:stab} briefly reviews the adopted method for stability analysis. 

\section{The reduced gyrofluid model} \label{sec:model}

Our analysis is based on a reduced nonlinear two-field gyrofluid model for collisionless  plasmas, which assumes the presence of a strong component of the magnetic field (strong guide field assumption) along one direction. The model consists of the following two evolution equations
\begin{align}
&  \frac{\partial N_e}{\partial t}+[\phi - \bpar, N_e]  - [\apar , U_e]=0,  \label{eq1red}\\
&  \frac{\partial \apar}{\partial t} + [\phi - \bpar, \apar  ] + \frac{1}{\Theta_e}[\apar,N_e]=0,  \label{eq2red}
\end{align}
complemented by the static relations 
\begin{align}
   &  N_e + (1 - \Gamma_{0i} + \Gamma_{1i})B_\pa + (1- \Gamma_{0i}) \frac{\phi}{\tau_{\pe _i}} =  0, \label{qnred}\\
   &  U_e=\bstar \lapp \apar     \label{ampparred} \\
   &  B_{\parallel}= - \frac{\bpe}{2}(N_e -(1 - \Gamma_{0i} + \Gamma_{1i})\phi + (1+ 2 \tau_{\perp_i}(\Gamma_{0i} - \Gamma_{1i}))B_{\parallel}).  \label{ampperpred}
\end{align}
which permit to express $\phi$, $U_e$  and $\bpar$ in terms of the dynamical variables $N_e$ and $\apar$.

Three independent  parameters are present in the system, and are given by
\begin{equation}   \label{param}
  \beta_{\pe_e}=8\pi \frac{ n_0 T_{0_{\perp e}}}{B_0^2}, \qquad   \tau_{\pe_i} = \frac{T_{0_{\perp  i}}}{T_{0_{\perp e}}},   \qquad   \Theta_e=\frac{T_{0_{\perp e}}}{T_{0_{ \parallel  e}}},  
\end{equation}
representing  the ratio between perpendicular electron pressure and guide field magnetic pressure, the ion-to-electron perpendicular temperature ratio, and the electron temperature anisotropy, respectively, at equilibrium.  In Eq. (\ref{param}) $B_0$ is the amplitude of the guide field, $n_0$ is the homogeneous equilibrium particle density (equal for both electrons and ions), $T_{0_{\perp e}}$ and $T_{0_{\perp i}}$ are the equilibrium temperatures in the plane perpendicular to the guide field for electrons and ions, respectively, whereas $T_{ {0_{ \parallel  e}}}$ is the electron equilibrium temperature along the direction of the guide field.

Equation (\ref{eq1red}) is the continuity equation for electron gyrocenters, whereas Eq. (\ref{eq2red}) is the component of a generalized Ohm's law along the direction of the guide field (which will be referred to as parallel direction, in the following, as opposed to "perpendicular", which, as customary, refers to the plane perpendicular to the guide field). 

 Eqs. (\ref{qnred}), (\ref{ampparred}) and (\ref{ampperpred}), on the other hand, correspond to the quasi-neutrality relation and to the parallel and perpendicular components of Amp\`ere's law, respectively, expressed in terms of gyrofluid variables. 

Adopting a Cartesian reference frame with coordinates $x$, $y$ and $z$, the fields $N_e$, $\apar$, $\phi$, $U_e$  and $\bpar$ are all functions of $x$ and $y$, as well as of the time coordinate $t$. The fields are defined over the two-dimensional domain  $\dn=\{(x,y) \in \mathbb{R}^2 \  |  \ 0 \leq x \leq 2\pi n , -L_y \leq y \leq L_y \}$, where $L_y$ is a constant and $n$ a non-negative integer. The choice of the 2D limit is motivated, on one hand, by the fact that, in such limit, the application of the Energy-Casimir method  becomes particularly fruitful, due to the abundance of Casimir invariants. Furthermore, chains of magnetic vortices observed, for instance in the solar wind, appear to have an essentially 2D structure \cite{perrone2017}. 

Periodic boundary conditions are assumed on the domain $\dn$. This is required in order for the gyroaverage operators $\Gamma_{0i}$ and $\Gamma_{1i}$ to be properly defined. Nevertheless, in the actual stability analysis, where two particular limits will be considered, a different domain will be adopted, as discussed in Sec. \ref{ssec:domain}.

In the evolution equations (\ref{eq1red})-(\ref{eq2red}), the symbol $[ \, , \, ]$ is defined by
\beq
[f,g]=\frac{\partial f}{\partial x}\frac{\partial g}{\partial y}-\frac{\partial g}{\partial x}\frac{\partial f}{\partial y},
\eeq
for two functions $f$ and $g$, and can be seen as a Poisson bracket with $x$ and $y$ as canonically conjugate variables. Such canonical bracket occurs very frequently in two-dimensional fluid models and is crucial for the existence of a Hamiltonian structure of Lie-Poisson type for such models (see, for instance Ref. \cite{Mor98}).

As above anticipated, $N_e$ and $\apar$ can be taken as the two dynamical variables of the system and represent the electron gyrocenter fluctuations and the parallel component of the magnetic vector potential (also referred to as magnetic flux function), respectively. The fields $\phi$, $U_e$ and $\bpar$, on the other hand, indicate the fluctuations of the electrostatic potential, of the parallel electron gyrocenter parallel velocity and of the parallel magnetic field, respectively.   
All variables are expressed in a normalized form, in the following way:
\begin{equation} \label{def}
\begin{split}
  &  x=\frac{\Tilde{x}}{\rhos}, \qquad    y=\frac{\Tilde{y}}{\rhos},  \qquad  t=\frac{\rhos}{\lpar}\omega_{ci} \Tilde{t},  \\
  & N_e=\frac{\lpar}{\rhos}\frac{\widetilde{N}_e}{n_0},   \qquad \apar=\frac{\lpar}{\rhos}\frac{\wapar}{B_0 \rhos}    \\
   & \phi=\frac{\lpar}{\rhos}\frac{e \wphi}{T_{0_{\perp e}}}, \qquad U_e=\frac{\lpar}{\rhos}\frac{\widetilde{U}_e}{c_{s \perp}}, \qquad   \bpar=\frac{\lpar}{\rhos}\frac{\wbpar}{B_0},\\
   \end{split}  
\end{equation}
where the tilde denotes the dimensional quantities. In Eq. (\ref{def})  $e$ is the proton charge and $\lpar$ is the characteristic length of the field fluctuations along the direction of the guide field (see \ref{sec:gyrokin} for the reduction of the original 3D model to 2D). Denoting with $m_i$ the mass of the ions present in the plasma and with $c$\ the speed of light, we also made use of the quantities $\omega_{ci}=e B_0/ (m_i c)$ indicating the ion cyclotron frequency, $c_{s \perp} =\sqrt{T_{0_{\perp e}} / m_i} $ indicating the sound speed based on the perpendicular temperature and $\rho_{s \perp}=c_{s \perp}/\omega_{ci}$, which is the sonic Larmor radius, also based on the perpendicular temperature. 

 We also introduced the short-hand notation $\bstar$ defined by
\begin{equation}   \label{bstar}
\bstar=\frac{2}{\bpe} + 1 - \frac{1}{\Theta_e},
\end{equation}
to indicate the modification due to electron temperature anisotropy in the parallel Amp\`ere's law (\ref{ampparred}). Note that $\bstar=2/\bpe$ in the isotropic case. We indicated with $\lapp$ the Laplacian operator in the perpendicular plane, so that $\lapp f =\partial_{xx} f +\partial_{yy} f$ for a function $f$.

The operators $\Gamma_{0i}$ and $\Gamma_{1i}$ represent the standard operators (see, e.g. Ref. \cite{Bri92}) associated with ion gyroaverage. We can define them in the following way. Let us consider a function $f=f(x,y)$, periodic over $\dn$ and indicate with $\scrd_n$ the lattice $\scrd_n=\{( l / n,  \pi m/  L_y ) : (l,m) \in \mathbb{Z}^2 \}$. We write the Fourier representation of $f$ as $f(x,y)=\sum _{\bk \in \scrd_n}  f_{\bk} \exp(i \bk \cdot \bx)$, where $\bx$ and $\bk$ are vectors of components $(x,y)$ and $(k_x, k_y )$, respectively, with $k_x =l/n$, $k_y=m \pi/L_y$, for $(l,m) \in \mathbb{Z}^2$. It is also convenient to introduce the quantity $b_i= \tpi k_\perp^2 $, where $k_\perp=\sqrt{k_x^2 + k_y^2}$, in adimensional variables, is the perpendicular wave number (in dimensional variables one would have $b_i=\tilde{k}_\perp^2 \rthi^2$ where $\tilde{k}_\perp$ is the dimensional perpendicular wave number and $\rthi=\sqrt{T_{0_{\perp  i}}/m_i}/\omega_{ci}$ is the perpendicular thermal ion gyroradius). The action of the operators $\Gamma_{0i}$ and $\Gamma_{1i}$ on the function $f$ is defined by
\begin{align}
& \Gamma_{0i} f (x,y)=\sum _{\bk \in \scrd_n} I_0 (b_i) \mathrm{e}^{-b_i} f_{\bk} \mathrm{e}^{i \bk \cdot \bx},  \label{gamma0}\\
& \Gamma_{1i} f (x,y)=\sum _{\bk \in \scrd_n} I_1 (b_i) \mathrm{e}^{-b_i} f_{\bk} \mathrm{e}^{i \bk \cdot \bx},   \label{gamma1}
\end{align}
with $I_0$ and $I_1$ indicating the modified Bessel functions of the first kind, of order $0$ and $1$, respectively.

In the light of the above definition, we can recognize in the first three terms of the continuity equation (\ref{eq1red}), the material derivative of the electron gyrocenter density, which is advected by a generalized incompressible velocity field $\mathbf{U}_{\perp e}=\hz\times \nabla (\phi - \bpar)$, where $\hz$ indicates the unit vector along the $z$ coordinate.  The latter originates from the electron gyrocenter velocity in the perpendicular plane, induced by electromagnetic perturbations. The last term, on the other hand, indicates the gradient of the parallel electron gyrocenter velocity along the perpendicular magnetic field. Equation (\ref{eq2red}) expresses, with its first two terms, the material derivative of the parallel vector potential $\apar$, advected by $\mathbf{U}_{\perp e}$. The third  term of the equation expresses the force exerted by the parallel gradient of the electron parallel pressure and is affected by temperature anisotropy. Alternatively, one could think at the terms $[\apar , N_e / \Theta_e +\bpar]$ as coming from the projection of the divergence of the anisotropic electron pressure tensor along the perpendicular magnetic field. The terms $\partial_t \apar +[\phi , \apar] $, on the other hand, come from the projection of the electric field along the magnetic field. In the quasi-neutrality relation (\ref{qnred}), the first two terms indicate the electron density fluctuations (recall that, in the limit of small electron FLR corrections, which is the case here, the relation $n_e=N_e + \bpar $ holds, where $n_e$ indicates the normalized electron density fluctuations, which differ from the electron gyrocenter density fluctuations  $N_e$ \cite{Bri92}). The remaining terms account for the contributions due to electromagnetic perturbations, and depend on ion finite Larmor radius effects, that arise when expressing the ion density fluctuations in terms of ion gyrocenter variables. Likewise, analogous contributions appear in Eq. (\ref{ampperpred}), upon replacing $N_e$ with $n_e - \bpar $. Parallel Amp\`ere's law (\ref{ampparred}) expresses the  fact that the parallel current density is proportional to the parallel electron gyrocenter velocity, the contribution of the gyrocenter ion velocity being negligible in the present model. This relation is also affected by the temperature anisotropy.

The present model shares some similarities with the model derived in Ref. \cite{Pas18}. More precisely, the two models coincide in the 2D limit if $\delta=0$ and $\Theta_e=1$.  Also,  if the pressure gradient term on the right-hand side of Eq. (\ref{eq2red}) is neglected, in the limit of isotropic electron temperature $\Theta_e=1$,  for cold ions ($\tpi \ll 1$), and with $\beta_{\perp e}$ small enough to neglect parallel perturbations $\bpar$, the system corresponds to 2D low-$\beta$ reduced MHD \cite{Str76,Kad74}. 
The derivation of the model, as well as its relation with similar models existing in the literature, are presented in  \ref{sec:gyrokin}.

\subsection{Hamiltonian structure}  \label{ssec:hamstruct}

The system (\ref{eq1red})-(\ref{ampperpred}) possesses a noncanonical Hamiltonian structure consisting of the Hamiltonian functional
\beq
H(N_e , \apar)= \frac{1}{2} \int_{\dn}  d^2 x \, \left( \frac{N_e^2}{\Theta_e} + b_{\star}|\nabla_{\perp}A_\pa|^2 -N_e \lphi N_e +N_e \lb N_e  \right),
    \label{hamred}
\eeq
and of the noncanonical Poisson bracket
\begin{equation}
\begin{split}
     & \{ F , G \}= \int_{\dn}  d^2 x \, ( N_e  [F_{N_e}, G_{N_e}]   + \apar([F_{\apar} , G_{N_e}] + [F_{N_e} , G_{\apar}])  ).
    \label{pbred}
\end{split}
\end{equation}
The evolution equations (\ref{eq1red}) and (\ref{eq2red}) can thus be written in the Hamiltonian form 
\beq
\frac{\partial \apar}{\partial t} = \{\apar, H \}, \qquad \, \frac{\partial N_e}{\partial t} = \{N_e, H\},
\eeq
with $H$ and $\{ \, , \, \}$ defined in Eqs. (\ref{hamred}) and (\ref{pbred}), respectively.

In Eq. (\ref{hamred}) we indicated with $\lphi$ and $\lb$ the linear operators that permit to express $\phi$ and $\bpar$ in terms of the dynamical variable $N_e$, so that
\beq
\phi = \lphi N_e, \qquad \bpar = \lb N_e.
\eeq
In Fourier space such operators take the form of multiplication operators and their explicit form can be found by solving the inhomogeneous linear system consisting of Eqs. (\ref{qnred}) and (\ref{ampperpred}), in the unknowns $\phi$ and $\bpar$ (see also Ref. \cite{Tas19}). Using the fact that, for $b_i >0$, one has $1- \Gamma_{0i}(b_i)>0$ and $\Gamma_{0i} (b_i) - \Gamma_{1i}(b_i)>0$, it is straightforward to see that, for Fourier modes with $b_i>0$, the system can always be solved and admits one solution. The resulting operators $\lphi$ and $\lb$ are symmetric with respect to the inner product $< f \, | \, g>=\int_{\dn} d^2 x \, fg$. This permits to verify that the functional $H$ is a conserved quantity (for finite $\tau_{\perp_i}$ and $b_i=0$, corresponding to the mode $(k_x,k_y)=(0,0)$,  one can fix equal to zero the corresponding Fourier coefficients, i.e. $B_{\parallel_{0,0}}=\phi_{0,0}=0$).
 
 In Eq. (\ref{pbred}), we introduced the notation $F_{f} = \delta F /\delta f$ to indicate the functional derivative of a functional $F$ with respect to a function $f$. The validity of the bilinear operator (\ref{pbred}) as Poisson bracket (and in particular the property of satisfying the Jacobi identity) is inherited from that of a more general Hamiltonian model, as discussed in \ref{sec:gyrokin}. Otherwise, in a more straightforward way, one can note that the bracket (\ref{pbred}) corresponds to the Poisson bracket of 2D reduced MHD \cite{Mor84}. As a consequence, Eqs. (\ref{eq1red})-(\ref{eq2red}), in addition to the Hamiltonian (\ref{hamred}), possess infinite conserved functionals, given by
\begin{equation}
    C_{1}= \int_{\dn} d^2x \, N_e \F(A_\pa),  \qquad C_{2}= \int_{\dn} d^2x \, \G(A_\pa),
    \label{eq:c1}
\end{equation}
where $\F$ and $\G$ are arbitrary functions. The functionals $C_1$ and $C_2$ are Casimir invariants of the Poisson bracket (\ref{pbred}) and, as such, they satisfy $\{C_1, E \}=\{C_2, E \}=0$ for every functional $E$. As discussed in Ref. \cite{Mor84}, the Casimir invariant $C_1$ includes, among others, the conservation of the integral of $N_e$ over an area bounded by contour lines of $\apar$. The Casimir $C_2$, on the other hand, expresses, for $\G(\apar)=\apar$, conservation of magnetic helicity at leading order.

\vspace{0.8cm}

\subsection{The domain of analysis}  \label{ssec:domain}

We intend to analyse the linear stability of equilibria such that the equilibrium solution for the magnetic flux function $\apar$ corresponds to the "cat's eyes" solution

\beq  \label{ceyes}
A_{eq}(x,y)=-\log (a \cosh y + \sqrt{a^2 -1} \cos x),
\eeq
where $a > 1$.

\begin{figure}
\centering
\includegraphics[width=10cm]{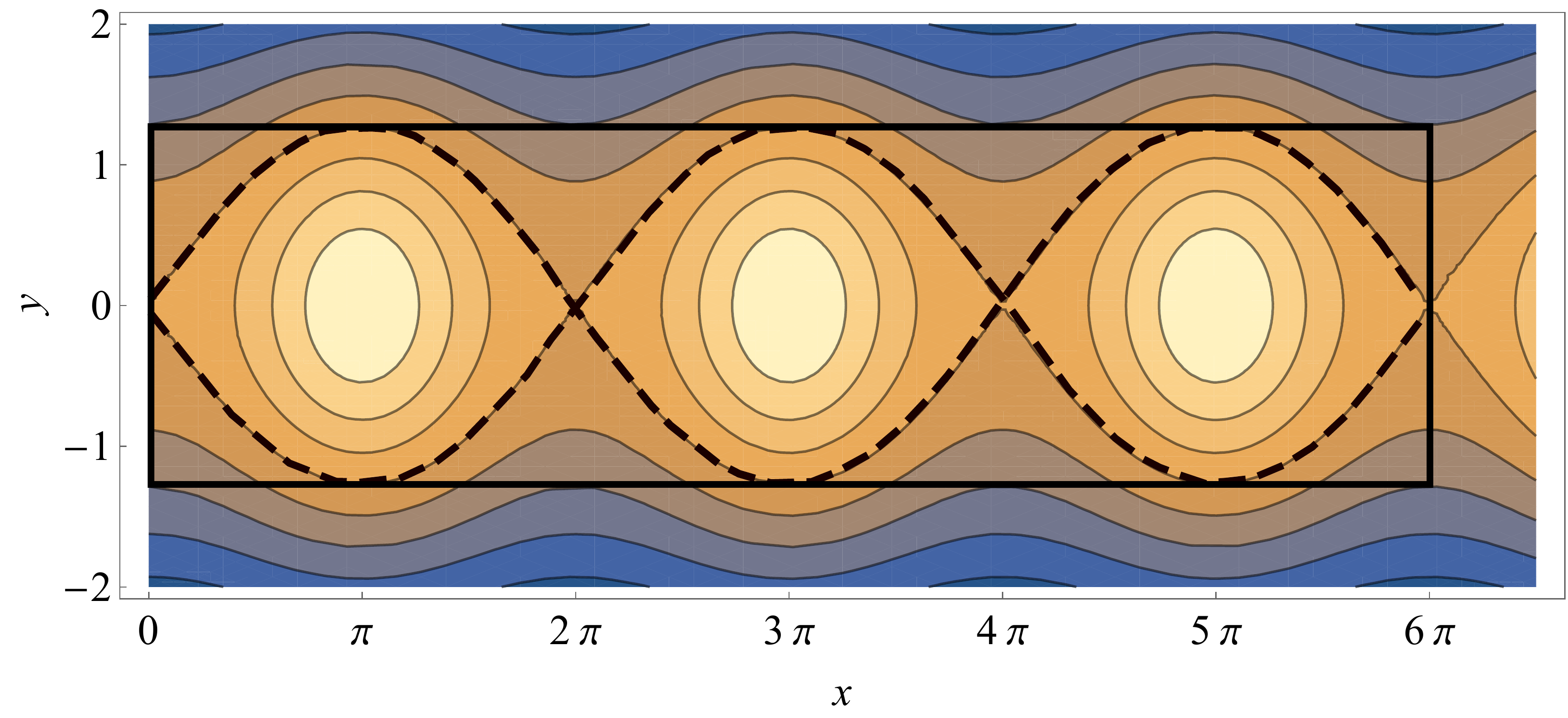}
\caption{The figure shows a surface plot and some contour lines of the "cat's eyes" function $A_{eq}$. The domain $D_n$, enclosed by the  separatrices, indicated with black dotted curves, and the domain $R_n$, corresponding to the rectangle enclosed by black solid lines, are also shown. The figure refers to the case $n=3$ and $a=1.12$. }
\label{fig1}
\end{figure}

As shown in Fig. \ref{fig1}, the contour lines of the function $A_{eq}$, in general describe chains of magnetic vortices in the plane $xy$. As $a \rightarrow 1^+$, the equilibrium configuration tends toward a uni-directional sheared magnetic field, with no magnetic vortices. 

The function $A_{eq}$ is known to be a solution of the Liouville's equation
\beq  \label{liouv}
\lapp \apar = - \ee^{2\apar}.
\eeq
Inspired by the procedure followed in Ref. \cite{Hol86b}, we carry out the stability analysis of the magnetic vortex chains on the domain
\begin{equation}  \label{domain}
    D_n= \Big\{ (x,y) \in \mathbb{R}^2 \    |  \   0 \leq x \leq 2  \pi n, \ \   |y| \leq \cosh^{-1}\Big( 1 + \small{\frac{\sqrt{a^2 -1}}{a}}\big(1 - \cos x \big)\Big)\Big\},
\end{equation}
with $a>1$. 

An example of such domain is depicted in Fig. \ref{fig1}. One can see that the domain corresponds to the domain bounded by the separatrices of the vortex chain and the number $n$ indicates the number of vortices in the domain. This choice for the domain allows, with appropriate boundary conditions, for the application of the Poincar\'e  inequality, which is crucial for carrying out some stability estimates that will be required later in the analysis. Of course, the choice of such domain rules out the effect of perturbations coming from outside the vortex chain. This can indeed be seen as a limitation of the present analysis. However, numerical simulations \cite{Puc18, Gra12,Gra09} show that, for instance, secondary instabilities due to colliding jets, can originate inside a magnetic island, and the subsequent turbulent evolution of the instability remains confined within the island. Therefore, in addition to the above mentioned technical argument related to the Poincar\'e inequality, restricting the analysis to the region enclosed by the separatrices does not appear to rule out all physically relevant processes. On the other hand, as pointed out in Ref. \cite{Hol86b}, this prevents from a direct comparison with classical results on stability of magnetic island chains, such as those of Refs. \cite{Fin77,Pri79,Bon83b}.

The domain $D_n$ differs from the domain $\dn$ introduced in Sec. \ref{sec:model} and different boundary conditions will have to be adopted. In particular, the definitions (\ref{gamma0}) and (\ref{gamma1}) of ion gyroaverage operators $\Gamma_{0i}$ and $\Gamma_{1i}$, are valid for a periodic domain. Variants of gyroaverage operators, that permit to account for different boundary conditions (e.g. Dirichlet), have been discussed, for instance in Refs. \cite{Gra06,Cro10}. These variants are often based on Taylor expansions or Pad\'e approximants, and on the identification between the quantity $b_i$ and the operator $-\tpi \lapp$. We follow the same practice and, in the two cases that will be treated in the subsequent Sections, we will use (very rough) approximants of the ion gyroaverage operators in two opposite limits.

\section{Hot-ion case : $\tpi \gg 1$}  \label{sec:hions}

We consider the model (\ref{eq1red})-(\ref{ampperpred}) in the limit $\tpi \gg1$, corresponding to an equilibrium perpendicular ion temperature much larger than the corresponding electron temperature. This limit was adopted for instance in the model of Ref. \cite{Che17} to describe turbulence at kinetic scales in the magnetosheath. Because of the relation $\rthi= \sqrt{\tpi}\rhos$, considering the limit $\tpi \gg 1$ implies that the characteristic length of our model, corresponding to $\rhos$, is much smaller than the perpendicular thermal ion gyroradius $\rthi$. Therefore, in this sense, one can refer to this limit, also as to a sub-ion limit. 
Because $I_0(b_i)\ee^{-b_i}=I_0(\tpi k_\perp^2) \ee^{-\tpi k_\perp^2}\rightarrow 0$, as $\tpi \rightarrow + \infty$, for all $(k_x,k_y)\in \scrd_n\setminus (0,0)$ and $I_1(b_i)\ee^{-b_i}=I_1(\tpi k_\perp^2) \ee^{-\tpi k_\perp^2}\rightarrow 0$, as $\tpi \rightarrow + \infty$, for all $(k_x,k_y)\in \scrd_n$ we simply take, for $\tpi \gg 1$, the following approximated form for the operators $\Gamma_{0i}$ and $\Gamma_{1i}$ :
\beq
\Gamma_{0i} f(x,y)=0, \qquad \Gamma_{1i} f(x,y)=0,  \label{gammahi}
\eeq
for a function $f(x,y)$ with $(x,y)\in D_n$ (for the mode $(k_x , k_y)=(0,0)$, an agreement between the exact form  for $\Gamma_{0i}$ acting on functions over $\mathcal{D}_n$ and the approximated form, written in Eq. (\ref{gammahi}), for $\tpi \gg 1 $ and for functions over $D_n$ can be obtained if, in the former case, one restricts to functions $f(x,y,t)=\sum _{(k_x , k_y) \in \scrd_n}  f_{(k_x, k_y)}(t) \exp(i (k_x x +k_y y))$ such that $f_{(0,0)}=0$, i.e. functions with zero spatial average). 

In the limit $\tpi \gg 1$, the model (\ref{eq1red})-(\ref{ampperpred}) thus reduces to
\begin{align}
&\frac{\partial N_e}{\partial t}- \bstar [\apar, \lapp \apar]=0,  \label{e1hi}\\
&\frac{\partial \apar}{\partial t}-\kappa [N_e , \apar]=0,  \label{e2hi}
\end{align}
with 
\beq   \label{bphihi}
\bpar = -N_e, \qquad \phi=-\frac{2}{\bpe}N_e.
\eeq
In Eq. (\ref{e2hi}) we also introduced the parameter
\beq   \label{kappa}
\kappa=\frac{2}{\bpe}+\frac{1}{\Theta_e}-1.
\eeq
For $\Theta_e=1$, i.e. for isotropic temperature, the model is analogous to the 2D version of the reduced electron MHD model discussed in Refs. \cite{Sch09} and \cite{Bol13}.

Equations (\ref{e1hi})-(\ref{e2hi}) are supplemented with the boundary conditions
\begin{align}  
&\apar \vert_\bd = a_A,   \label{bchi1}\\
&  N_e \vert_\bd = a_N,  \label{bchi2}
\end{align}
with $a_A, a_N \in \mathbb{R}$ and where we indicated with $\bd$ the boundary of $D_n$. 
The boundary condition (\ref{bchi1}) expresses the fact that the perpendicular magnetic field $\mathbf{B}_\perp=\nabla \apar \times \hz$ is tangent to the boundary, i.e. $\mathbf{B}_\perp \cdot \bn =0$, where $\bn$ is the outward unit vector normal to the boundary $\bd$. The condition (\ref{bchi2}), on the other hand, implies $\mathbf{U}_{\perp e} \cdot \bn =0$, meaning that the incompressible flow $\mathbf{U}_{\perp e}=\hz\times\nabla (\phi - \bpar)=(1-2/ \bpe)\hz \times \nabla N_e$ is tangent to the boundary. 

The procedure we adopt to investigate the linear stability of magnetic vortex chains is summarized in  \ref{sec:stab}. Detailed descriptions of the method can be found in Refs. \cite{Mor98} and \cite{Hol85}.
The first step consists of finding a functional $F$ given by a combination of constants of motion of the system (\ref{e1hi})-(\ref{e2hi}).  To this purpose we can use the Hamiltonian (\ref{hamred}) and the Casimir invariants (\ref{eq:c1}), with $\phi$ and $\bpar$ given by Eq. (\ref{bphihi}). Indeed, such functionals are also conserved by the system  (\ref{e1hi})-(\ref{e2hi}) on the domain $D_n$. This can be shown by direct computation making use of the identities
\begin{align}
&\intd d^2 x \, f \lapp g=-\intd d^2 x \, \nabla f \cdot \nabla g + \intbd \, f \frac{\partial g}{\partial n} ds, \label{id1}\\
&\intd d^2 x \, f [g,h]=\intd d^2 x \, h[f,g]- \intbd \, hf \nabla g \cdot \dl, \label{id2}
\end{align}
for functions $f, g$ and $h$, and of the boundary conditions (\ref{bchi1}) and (\ref{bchi2}). In Eq. (\ref{id1}) we indicated with $ds$ the scalar infinitesimal arc element and with $\partial g / \partial n =\nabla g \cdot \bn$ the gradient normal to the boundary. In Eq. (\ref{id2}) we indicated with $\dl$ the vectorial infinitesimal arc element.

\subsection{First variation and equilibria}

We consider then the conserved functional $F=H+C_1+C_2$, explicitly given by
\beq
F(N_e , \apar)=\intd d^2x \, \left( \bstar \frac{\vert \nabla \apar \vert^2}{2}+ \kappa \frac{N_e^2}{2}+N_e \F (\apar) + \G (\apar)\right).
\eeq
Adopting, for the variations $\delta \apar$ and $\delta N_e$, the boundary conditions
\beq  \label{bcperthi}
\delta \apar \vert_\bd =0, \qquad \delta N_e \vert_\bd =0,
\eeq
the first variation of $F$ is given by
\begin{align}
& \delta F (N_e, \apar ; \delta N_e , \delta \apar)= \\ \nonumber
&\intd d^2x \, \left( (- \bstar \lapp \apar + \F ' (\apar ) N_e + \G ' (\apar ))\delta \apar + (\kappa N_e +\F (\apar ))\delta N_e\right),
\end{align}
where the prime denotes derivative with respect to the argument of the function.

Setting the first variation $\delta F$ equal to zero for arbitrary perturbations, leads to the system
\begin{align}
& \lapp \apar =\frac{\F ' (\apar ) N_e}{\bstar} + \frac{\G ' (\apar)}{\bstar}, \label{eq1hi} \\
&\F (\apar )= - \kappa N_e,   \label{eq2hi}
\end{align}
Solutions of Eqs. (\ref{eq1hi})-(\ref{eq2hi}) are equilibrium solutions of the system (\ref{e1hi})-(\ref{e2hi}). Eq. (\ref{eq1hi}) can be seen as a Grad-Shafranov equation for the current density $-\lapp \apar$, whereas Eq. (\ref{eq2hi}) expresses the fact that the electron gyrocenter density fluctuations $N_e$ (and, by virtue of Eq. (\ref{bphihi}), the electrostatic potential and the parallel magnetic perturbations) are constant on perpendicular magnetic field lines identified by $\apar= \mathrm{constant}$. For such equilibria $\F (\apar)=\kappa (\bpe /2) \phi$, and in particular $\F(\apar)=\phi$ for isotropic temperature. Therefore, for $\F=0$ we obtain an equilibrium with no perpendicular equilibrium flow. For $\F (\apar)=\pm \sqrt{2/ \bpe} \apar $ and assuming isotropic temperature, on the other hand, one obtains Alfv\'enic solutions, in which the equilibrium $\mathbf{E}\times\mathbf{B}$ velocity field, given by $\hz \times \nabla \phi$, equals, in dimensional units, the local Alfv\'en velocity field (or its opposite). In the more general case with $\Theta_e \neq 1$, the Alfv\'en velocity will be modified by an effect due to temperature anisotropy.  When $\F$ is taken as a linear function of $\apar$, clearly also the perpendicular equilibrium flow exhibits the "cat's eyes" pattern. 

The system is characterized by the two arbitrary functions $\F$ and $\G$. Because we are interested in solutions for $\apar$ given by the "cat's eyes" function (\ref{ceyes}), we constrain Eq. (\ref{eq1hi}) to equal the Liouville equation (\ref{liouv}) (we consider here non-propagating solutions but a generalization to account for a constant propagation velocity could be carried out). This occurs if the following condition on the function $\G$ is fulfilled:
\beq  \label{ghi}
\G (\apar)=-\frac{\bstar}{2}\ee^{2\apar}+\frac{\F^2 (\apar)}{2 \kappa} + c_1,
\eeq
with $c_1$  arbitrary constant.

Our analysis will then focus on the class of equilibria given by
\begin{align}
&\apar =A_{eq},   \label{equi1hi}\\
&N_e = -\frac{\F (A_{eq})}{\kappa},  \label{equi2hi}
\end{align}
for $\kappa \neq 0$, with $A_{eq}$ given by Eq. (\ref{ceyes}) and arbitrary $\F$. The corresponding expressions for $\phi$ and $\bpar$ at equilibrium are given by $\phi=2\F (A_{eq})/(\bpe \kappa)$ and $\bpar=\F (A_{eq})/ \kappa$, respectively. Therefore, we note that, for $\tpi \gg 1$, equilibria obtained from the above variational principle and possessing a magnetic vortex chain, admit a whole class of flows (or, equivalently, of electron gyrocenter density or parallel magnetic perturbations) depending on an arbitrary function. 
 
 \subsection{Second variation and stability conditions}

 The second variation of $F$, making use of the boundary conditions (\ref{bcperthi}) and rearranging terms, can be written as
 \begin{align}
& \delta^2 F(\apar , N_e ; \delta \apar , \delta N_e) =\intd  d^2 x \, \left(\bstar \vert \nabla \delta \apar \vert^2 +( \F '' (\apar) N_e + \G '' (\apar) - {\F'}^2 (\apar))\vert\delta \apar\vert^2 \right. \nonumber \\
& \left. + (\kappa -1)\vert \delta N_e\vert^2
 +(\F ' (\apar) \delta \apar  + \delta N_e)^2\right)
 \end{align}
We intend to find conditions for which $\delta^2F$, evaluated at the class of equilibrium of interest, is positive for arbitrary perturbations. If we impose $\bstar >0$ and $\kappa >1$, it is only the coefficient of $\vert \delta \apar\vert^2$ that can provide a negative contribution, and thus indefiniteness, to $\delta^2 F$. Using the relation (\ref{ghi}) one finds that, for the class of equilibria of interest, such coefficient is given by  $-2 \bstar \ee^{2 A_{eq}} +(1/ \kappa -1){\F '} ^2 (A_{eq})$. For $\kappa >1$ this coefficient is always negative, so the second variation has no definite sign. This indefiniteness seems to reflect a feature of "cat's eyes" equilibria that was already pointed out in Ref. \cite{Mar87} in the case of the 2D Euler equation for an incompressible flow.  This difficulty can be overcome, as indicated in Ref. \cite{Mar87}, by making use of a Poincar\'e inequality. In our specific case, the required Poincar\'e inequality reads
\beq  \label{poinc}
\intd  d^2 x \, \vert \nabla \delta \apar \vert^2 \geq \km \intd d^2 x \, \vert \delta \apar \vert^2,
\eeq
with $\delta \apar \vert_\bd =0$.  In the inequality (\ref{poinc}), $\km$ is the minimal eigenvalue of the operator $-\lapp$ acting on the functions defined over $D_n$ and vanishing on the boundary of $D_n$. The inequality (\ref{poinc}) can be derived with a straightforward modification of the procedure followed in Ref. \cite{Hol86b}. Following this same Reference, we make use of the fact that  $\km > \kr$, where $\kr$ is the minimal eigenvalue of  the operator $-\lapp$ on the functions defined over $R_n$ and vanishing on the boundary of $R_n$. The domain $R_n \supset D_n$ is defined by
\beq  \label{Rn}
R_n= \Big\{ (x,y) \in \mathbb{R}^2\ \   |\ \ 0 \leq x \leq 2 n \pi, \ \  |y| \leq l = \cosh^{-1}\Big( 1 + 2 \small{\frac{\sqrt{a^2 -1}}{a}}\Big) \Big\}
\eeq
and corresponds to the rectangle of width $2n\pi$ and height $2l$ equal to the magnetic island width. The rectangle $R_n$ is depicted in Fig. \ref{fig1}. For perturbations vanishing on the boundary of $R_n$, one has 
\beq  \label{kr}
\kr=\frac{1}{4n^2}+\frac{\pi^2}{4 l^2}.
\eeq
With regard to this point, we remark that the expression for the minimal eigenvalue (\ref{kr}) differs by a factor $4$ from the one used in Ref. \cite{Hol86b} for the fluid case. The reason for this difference is due to the fact that in Ref. \cite{Hol86b}, in order to obtain the equilibrium equation, the perturbations of the stream function were assumed to vanish on the boundary and to have zero circulation along the boundary. In our case, in order to obtain the desired equilibrium equations for the magnetic field, it is sufficient to impose that the perturbations of $\apar $ and $N_e$ vanish on the boundary.

With the help of the above reasoning, we can state that, for $\bstar >0$
\begin{align}  \label{boundhi}
&\delta^2 F ( A_{eq} , \F (A_{eq}); \delta \apar, \delta N_e) \geq \intd  d^2 x \, \left(  \left( \bstar \kr  - 2 \bstar \ee^{2 A_{eq}} +(1/ \kappa -1){\F '} ^2 (A_{eq}) \right)\vert\delta \apar\vert^2 \right. \nonumber \\
& \left. + (\kappa -1)\vert \delta N_e\vert^2 +(\F ' (A_{eq}) \delta \apar  + \delta N_e)^2\right).
 \end{align}
 The coefficient of $\vert \delta \apar\vert^2$ on the right-hand side of Eq. (\ref{boundhi}) can be made positive by choosing appropriate bounds for ${\F '}^2 (A_{eq})$. In particular, noticing that
 \beq
 \min_{(x,y) \in D_n} (- 2 \bstar \ee^{2 A_{eq}(x,y)})=-2 \bstar \ee^{2 A_{eq}(\pi ,0)}=-\frac{2 \bstar}{(a - \sqrt{a^2 -1})^2},
 \eeq
 one can write that, for $(x,y) \in D_n$:
 \begin{align}  
 &\bstar \kr  - 2 \bstar \ee^{2 A_{eq}(x,y)} +\left(\frac{1}{ \kappa} -1 \right){\F '} ^2 (A_{eq}(x,y)) \nonumber \\
 & \geq \bstar \left( \kr -\frac{2 }{(a - \sqrt{a^2 -1})^2} \right)+\left(\frac{1}{ \kappa} -1 \right){\F '} ^2 (A_{eq}(x,y)).  \label{bound2hi}
 \end{align}
 Making use of the relations (\ref{boundhi}), (\ref{bound2hi}), (\ref{kr}) as well as of the previously mentioned conditions $\bstar >0$ and $\kappa >1$, we can conclude that the linear stability of the family of equilibria (\ref{equi1hi})-(\ref{equi2hi}) is attained if the following three conditions are satisfied:
 \begin{align}
 & \bstar >0, \label{cond1hi}\\
 &\kappa>1,  \label{cond2hi}\\
 &\bstar \left( \frac{1}{4n^2}+\frac{\pi^2}{4 l^2} -\frac{2 }{(a - \sqrt{a^2 -1})^2} \right) \geq \max_{(x,y)\in D_n} \left( 1 - \frac{1}{\kappa}\right){\F '} ^2 (A_{eq} (x,y)). \label{cond3hi}
 \end{align}
 Note that the right-hand side of Eq. (\ref{cond3hi}) is not negative when the condition (\ref{cond2hi}) is fulfilled.
 
 In order to get some physical insight from these conditions we resort first to the definitions (\ref{bstar}) and (\ref{kappa}). In terms of the perpendicular electron beta parameter $\bpe$ and on the electron temperature anisotropy parameter $\Theta_e$, the conditions (\ref{cond1hi}) and (\ref{cond2hi}) imply
 \begin{equation}  \label{cond1bhi}
 \Theta_e > \frac{\beta_{\pe _e}}{2 +\beta_{\pe _e}}, \qquad   \mbox{if $0 < \beta_{\pe _e} \leq 1$},
\end{equation}
\begin{equation}   \label{cond2bhi}
 \frac{\beta_{\pe _e}}{2 +\beta_{\pe _e}} < \Theta_e < \frac{\beta_{\pe _e}}{2(\beta_{\pe _e} - 1)}   \qquad \mbox{if $ 1 < \beta_{\pe _e} < 4$}.
\end{equation}
 From the relations (\ref{cond1bhi}) and (\ref{cond2bhi}) it emerges that the stability conditions imply an upper bound $\bpe=4$ for the perpendicular electron plasma beta parameter. This bound appears not to be too restrictive for typical solar wind or magnetospheric parameters.    We also observe that the condition $\Theta_e > \bpe/(2 + \bpe)$, that emerges in our analysis in both Eq. (\ref{cond1bhi}) and (\ref{cond2bhi}) (and which corresponds to $\bstar >0$), is the condition that suppresses the firehose instability in the stability analysis of spatially homogeneous equilibria based on linear waves (see, e.g. Ref. \cite{hasegawa}).  Although our conditions are sufficient but not necessary, we could argue that also magnetic vortex chains could be subject to the same instability.  For $1 < \bpe < 4$ an upper bound for temperature anisotropy also appears. This is due to the  condition (\ref{cond2hi}). However, unlike the lower bound, this bound does not appear to be related to instability thresholds familiar from wave linear theory and in particular to those, such as mirror instability (see, e.g. Ref. \cite{hasegawa}), occurring when the temperature anisotropy parameter $\Theta_e$ is too large.

The condition  (\ref{cond3hi}), on the other hand, involves directly the structure of the magnetic vortex chain and of the equilibrium electron gyrocenter density (or, equivalently, of the equilibrium electrostatic potential or of the parallel magnetic perturbations). Inserting the expression for the length $l$ in terms of $a$, which can be extracted from Eq. (\ref{Rn}), the condition (\ref{cond3hi}) can be reformulated as
\begin{align}
&\bstar\left(    \frac{1}{4n^2}+\frac{\pi^2} {4 \left(\cosh^{-1}\Big( 1 + 2 \small{\frac{\sqrt{a^2 -1}}{a}}\Big)\right)^2 }  -\frac{2 }{(a - \sqrt{a^2 -1})^2}        \right)    \nonumber \\
&\geq  \max_{(x,y)\in D_n} \left( 1 - \frac{1}{\kappa}\right){\F '} ^2 (A_{eq} (x,y)).   \label{cond3bhi}
 \end{align}
 Obviously, this condition depends on the choice of the arbitrary function $\F$.  For the choice $\F=0$, which corresponds to $\phi=0$ at equilibrium, and thus no perpendicular flow, the right-hand side of 
 Eq. (\ref{cond3bhi}) vanishes. If we consider a single vortex ($n=1$) in the absence of flow (i.e. the most favorable situation for stability), then, for $\bstar>0$, one can verify numerically that the condition (\ref{cond3bhi}) is satisfied for
 \beq
 1 < a < 1.026..
 \eeq
  From Eq. (\ref{cond3bhi}), it also transpires that considering longer chains of vortices by increasing $n$, makes it more difficult to satisfy the stability condition. For instance, for $n=4$, always in the absence of perpendicular flow, one has that the stability condition is satisfied for $1 < a < 1.023..$.Considering the expression (\ref{ceyes}), this implies a ratio $\sqrt{a^2 -1}/a$, between the amplitude of the vortices and that of the background sheared magnetic field, equal at most to approximately $0.21$, in order to fulfill the stability condition.  
  
  When $\F (\apar)$ is chosen as a linear function $\F(\apar)=V_1 \apar$, with constant $V_1$, the condition  Eq. (\ref{cond3bhi}) becomes
 \begin{align}
&\bstar\left(    \frac{1}{4n^2}+\frac{\pi^2} {4 \left(\cosh^{-1}\Big( 1 + 2 \small{\frac{\sqrt{a^2 -1}}{a}}\Big)\right)^2 }  -\frac{2 }{(a - \sqrt{a^2 -1})^2}        \right)    \nonumber \\
&\geq   \left( 1 - \frac{1}{\kappa}\right)V_1^2.   \label{cond3hilin}
 \end{align}
 Because, at equilibrium $\F (A_{eq})=\kappa (\bpe /2) \phi$, from interpreting $\phi$ as a stream function for the equilibrium flow, it follows  that $V_1^2$ is proportional to the ratio between the square of the amplitude of the equilibrium flow and that of the local Alfv\'en velocity. The condition (\ref{cond3hilin}) can then be seen as an upper bound on the speed of the equilibrium flow. This condition is similar to the sub-Alfv\'enic condition emerging from the Energy-Casimir method applied to other plasma models \cite{Haz84,Hol85,Tro15}.

\section{Cold-ion case : $\tpi \ll 1$}  \label{sec:cions} 

In this Section we consider the opposite limit, i.e. $\tpi \ll 1$. This limit is adopted mainly for laboratory plasmas \cite{Bol13}.  In terms of scales, it implies that the characteristic scale $\rhos$ is much larger than the perpendicular ion thermal gyroradius $\rthi$.  

Based on the relations $I_0(\tpi k_\perp^2) \ee^{-\tpi k_\perp^2} =1-\tpi k_\perp^2+\mathcal{O}(\tpi^2)$ and $I_1(\tpi k_\perp^2) \ee^{-\tpi k_\perp^2}\rightarrow 0$, as $\tpi \rightarrow  0$, for all $(k_x,k_y)\in \scrd_n$, we consider the following approximations for the ion gyroaverage operators for the cold-ion limit :
\beq
\Gamma_{0i} f(x,y)=(1+\tpi \lapp )f(x,y)+\mathcal{O}(\tpi^2), \qquad \Gamma_{1i} f(x,y)=0,  \label{gammaci}
\eeq
for $f$ defined over the domain $D_n$.
With this prescription, the model (\ref{eq1red})-(\ref{ampperpred}) in the cold-ion limit becomes
\begin{align}
&\frac{\partial N_e}{\partial t}+[\phi , N_e]- \bstar [\apar, \lapp \apar]=0,  \label{e1ci}\\
&\frac{\partial \apar}{\partial t}+[\phi , \apar]+\lambda [N_e , \apar]=0,  \label{e2ci}
\end{align}
with 
\beq   \label{bphici}
\bpar = -\frac{\bpe}{2 + \bpe}N_e, \qquad \lapp \phi=N_e
\eeq
and the parameter $\lambda$ defined by
\beq
\lambda=\frac{\bpe}{2 + \bpe}-\frac{1}{\Theta_e}.
\eeq
The parameter $\lambda$ is associated with the terms coming from the divergence of the anisotropic electron pressure tensor.
In the limit of isotropic temperature ($\Theta_e=1$) and when $\bpar$ is negligible, this model can be seen as the two-field model studied in Ref. \cite{Caf98} in the limit of vanishing electron inertia. If, furthermore, the third term on the left-hand of Eq. (\ref{e2ci}) is also neglected, the model becomes analogous to 2D low-$\beta$ reduced MHD.  

We adopt the following boundary conditions:
\begin{align}  
&\apar \vert_\bd = a_A,   \label{bcci1}\\
&  \phi \vert_\bd = a_\phi,  \label{bcci2}
\end{align}
with $a_A, a_\phi \in \mathbb{R}$. The boundary condition (\ref{bcci1}) is identical to Eq. (\ref{bchi1}) and implies $\mathbf{B}_\perp \cdot \bn=0$. Equation (\ref{bcci2}), analogously to Eq. (\ref{bchi2}), refers to a condition of a velocity field tangent to the boundary.  However, in the hot-ion case, because of the proportionality between $\phi$ and $\bpar$,  the condition applied to the entire field $\mathbf{U}_{\perp e}=\hz\times\nabla(\phi - \bpar)$. In the cold-ion case, $\phi$ and $\bpar$ are no longer proportional, so that the condition (\ref{bcci2}) expresses the fact that the normalized $\mathbf{E}\times\mathbf{B}$ velocity field, given by $\ueb=\hz \times \nabla \phi$, is tangent to the boundary, i.e. $\ueb \cdot \bn =0$.

With the help of the identities (\ref{id1})-(\ref{id2}) and applying the boundary conditions (\ref{bcci1})-(\ref{bcci2}), it is possible to show that the functionals given in (\ref{hamred}) and (\ref{eq:c1}), with $\bpar$ and $\phi$ related to $N_e$ by Eq. (\ref{bphici}), are conserved by the system (\ref{e1ci})-(\ref{e2ci}) on the domain  $D_n$. Therefore, we can consider the constant of motion $F=H+C_1+C_2$ given by
\beq   \label{fconsci}
F(N_e , \apar)=\intd d^2x \, \left( \bstar \frac{\vert \nabla \apar \vert^2}{2}+ \frac{\vert \nabla \phi \vert^2}{2}-\lambda \frac{N_e^2}{2}+N_e \F (\apar) + \G (\apar)\right).
\eeq
We remark that, although $F$ is a functional of $N_e$ and $\apar$, we also used, for convenience, the variable $\phi$ for its expression on the right-hand side of Eq. (\ref{fconsci}). We point out that $\phi$ has to be intended as the unique solution of the problem $\lapp \phi =N_e$, with $ \phi \vert_\bd = a_\phi$. In this way, the field $\phi$ can be interpreted as $\phi = \lapp^{-1} N_e$ and is unambiguously defined for a given $N_e$. 

\subsection{First variation and equilibria}  \label{ssec:fvarci}

We impose the following boundary conditions for the perturbations of $\apar$ and $\phi$:
\beq  \label{bcpertci}
\delta \apar \vert_\bd =0, \qquad \delta \phi \vert_\bd =0, \qquad \intbd \frac{\partial \delta \phi}{\partial n} ds =0.
\eeq
Analogously to the case of the field $\phi$, also the perturbation $\delta \phi$ has to be interpreted as the solution of the problem $\lapp \delta \phi =\delta N_e$, with $ \delta \phi \vert_\bd = 0$, with $\delta N_e$ indicating the perturbation of the dynamical variable $N_e$. The two boundary conditions concerning $\delta \phi$ correspond to those also adopted in Ref. \cite{Hol86b}. Indeed, in the cold-ion case, the second term on the right-hand side of Eq. (\ref{fconsci}) is analogous to the kinetic energy term in the conserved functional of the 2D Euler equation for an incompressible fluid.

Subject to the boundary conditions (\ref{bcpertci}), the first variation of $F$ reads
\begin{align}
& \delta F (N_e, \apar ; \delta N_e , \delta \apar)= \\ \nonumber
&\intd d^2x \, \left( (- \bstar \lapp \apar + \F ' (\apar ) N_e + \G ' (\apar ))\delta \apar + (\F (\apar )-\lambda N_e - \phi)\delta N_e\right).
\end{align}
Setting the first variation equal to zero leads to the following equilibrium equations:
\begin{align}
& \lapp \apar =\frac{\F ' (\apar ) N_e}{\bstar} + \frac{\G ' (\apar)}{\bstar}, \label{eq1ci} \\
&\F (\apar )= \phi +\lambda N_e,   \label{eq2ci}
\end{align}
Imposing that $\apar$ satisfies Liouville's equation implies that Eq. (\ref{eq1ci}) becomes
\beq  \label{eq1cibis}
\G ' (\apar)=-N_e \F ' (\apar)-\bstar \ee^{2 \apar}.
\eeq
We consider first the case where $\F ' (\apar) \neq 0$.  In this case, from Eq. (\ref{eq1cibis}), one has
\beq
N_e=-\frac{\bstar \ee^{2 \apar} + \G ' (\apar)}{\F ' (\apar)},
\eeq
from which it follows that, at equilibrium, $N_e=N_e (\apar)$. Equation (\ref{eq2ci}) thus implies that also $\phi=\phi (\apar)$, for the equilibria of interest. Using this fact in Eq. (\ref{eq1cibis}), together with the relation $\lapp \phi = N_e$, leads to the equation
\beq  \label{releqci}
\phi '' (\apar) \vert \nabla \apar \vert^2=-\frac{\G ' (\apar)}{\F ' (\apar)}+\frac{\ee^{2 \apar}}{\F ' (\apar)} (\phi ' (\apar ) \F ' (\apar )- \bstar).
\eeq
We specialize now to the solution of interest $\apar =A_{eq}$. Because the right-hand side of Eq. (\ref{releqci}) is a function of $\apar$ only, so has to be the left-hand side. In particular, for $\apar = A_{eq}$ one has to verify if $\vert \nabla A_{eq}\vert^2$ is a function of $A_{eq}$ only. In order to test this, we consider the function
\beq
\Upsilon (x,y)= \vert \nabla A_{eq}(x,y)\vert^2=\frac{(a^2 -1)\sin^2 x +a^2 \sinh^2 y}{(a \cosh^2 y + \sqrt{a^2 -1}\cos x )^2}.
\eeq
If $\vert \nabla A_{eq}\vert^2$ were a function of $A_{eq}$ only, then, upon the local change of coordinates $(x,y) \leftrightarrow (x ' , A_{eq})$ given by 
\begin{align}
& x = x',\\
& y = \cosh^{-1}\left(\frac{\ee^{-A_{eq}}}{a}-\frac{\sqrt{a^2 -1}}{a} \cos x' \right),
\end{align}
(invertible, for instance, for $0 < x < \pi$ and $0 < y < \cosh^{-1}(1+(\sqrt{a^2 -1}/a)(1-\cos x))$)
one would have $\Upsilon(x,y)=\bar{\Upsilon} (x' , A_{eq})=\bar{\Upsilon}(A_{eq})$, for every $x'$ in the domain of invertibility. However,
\beq
\Upsilon(x,y)=\bar{\Upsilon}(x' , A_{eq})=1 - \ee^{2 A_{eq}}-2 \sqrt{a^2 -1} \ee^{A_{eq}} \cos x'.
\eeq
Because $\partial \bar{\Upsilon} /\partial x' = 2 \sqrt{a^2 -1}\exp(A_{eq})\sin x' \neq 0$ (for instance for $0 < x' < \pi$), we conclude that $\bar{\Upsilon}$ is not constant with respect to $x'$ and thus $\vert \nabla A_{eq}\vert^2$ is not a function of $A_{eq}$ only on $D_n$. As a consequence, in order for Eq. (\ref{releqci}) to hold for "cat's eyes" equilibria, one has to set $\phi '' (A_{eq})=0$, which implies
\beq   \label{phieqci}
\phi=K_1 A_{eq} +K_2,
\eeq
with $K_1$ and $K_2$ arbitrary constants. As a consequence, using $N_e =\lapp \phi$, we obtain that the equilibria supporting magnetic vortex chains are given by
\begin{align}
&\apar=A_{eq}, \label{equi1ci} \\
&N_e= K_1 \lapp A_{eq} =-\frac{K_1}{(a \cosh y+\sqrt{a^2 -1} \cos x)^2},  \label{equi2ci}
\end{align}
with $K_1 \neq 0$.
From Eqs. (\ref{eq2ci}) and (\ref{eq1cibis}) one obtains that the corresponding choice for the arbitrary functions $\F$ and $\G$ are given by
 \begin{align}
 &\F(\apar)=-\lambda K_1 \ee^{2 \apar}+K_1 \apar +K_2,   \label{Ff1}\\
 &\G (\apar)=-\frac{\lambda K_1^2}{2}\ee^{4\apar}+\frac{K_1^2 - \bstar}{2}\ee^{2 \apar}+G_1,  \label{Gf1}
 \end{align}  
 with arbitrary constant $G_1$. 
 
 We recall that, in the case $\lambda=0$,  the problem of determining equilibrium solutions with flow can be circumvented \cite{Mor86b,Thr99}, in the case of sub-Alfv\'enic flows, by rewriting Eq. (\ref{eq1ci}) in terms of the new variable
 \beq
 u(\apar)=\int_0^{\apar} dg \,  \sqrt{1-{\F '}^2 (g)/ \bstar}.
 \eeq
 This transformation leads to a Grad-Shafranov equation (i.e. without flow) for the independent variable $u$. Once solutions for this equation are found, the corresponding equilibrium magnetic and velocity fields can be constructed. This procedure was applied also in Ref. \cite{Thr09}. However, it was applied to a magnetic field different from the one we obtain from Eq. (\ref{ceyes}), although it shares the same magnetic surfaces.
 
 The expressions for $\phi$ and $\bpar$ at equilibrium, on the other hand, are given by Eq. (\ref{phieqci}) and by $\bpar=-K_1( \bpe / (2 +\bpe) )\lapp A_{eq}$, respectively. For these equilibria, the $\ueb$ velocity is locally proportional to the perpendicular Alfv\'en velocity. The corresponding streamlines, therefore, exhibit the same pattern of the magnetic vortex chain. The electron gyrocenter density and the parallel magnetic perturbations, on the other hand, are proportional to the equilibrium current density given by $-\lapp A_{eq}$.
 
 In the case $\F ' (\apar)=0$ the equilibrium equations (\ref{eq1ci})-(\ref{eq2ci}) decouple. The "cat's eyes" solutions for the magnetic flux function are obtained with the choice
 \beq  \label{Gcib}
 \G ' (\apar)=- \bstar \ee^{2 \apar}.
 \eeq
 On the other hand, given that $\F ' (\apar)=0$ implies $\F (\apar)=F_1$, with $F_1$ arbitrary constant, Eq. (\ref{eq2ci}) yields
 \beq  \label{eqphi}
 \lambda \lapp \phi + \phi = F_1.
 \eeq
 Therefore, in this case, $\phi$ and $N_e$ are not constrained to be constant, at equilibrium, on the contour lines of $A_{eq}$, as in the previous case.
 We remark that in this case, unlike low-$\beta$ reduced MHD (formally retrieved by setting $\lambda=0$ and $\bstar=2/\bpe$), the choice $\F ' (\apar)=0$ does not necessarily lead to zero $\mathbf{E}\times\mathbf{B}$ flow. Indeed, the presence of the additional contribution due to the first term on the left-hand side of Eq. (\ref{eqphi}), originated from the electron pressure tensor, makes it possible to obtain non trivial flows in the presence of magnetic vortex chains.
 
 The equilibria considered in this case are thus given by
 \begin{align}
 &\apar = A_{eq},  \label{eq1cib} \\
 &N_e = \lapp \phi_{eq},  \label{eq2cib}
  \end{align}
  where $\phi_{eq}$ is a solution of Eq. (\ref{eqphi}) with boundary condition (\ref{bcci2}). Clearly, one can transform this problem into an equivalent problem for a homogeneous equation with Dirichlet boundary conditions, by introducing the new variable $\bar{\phi}=\phi - F_1$  and imposing the boundary condition $\bar{\phi}\vert_\bd=a_{\phi}-F_1$. Analytical solutions of this problem can be sought for, for instance with the method described in Ref. \cite{Rea96}.

 \subsection{Second variation and stability conditions}

The second variation of the functional (\ref{fconsci}) reads
 \begin{align}
& \delta^2 F(\apar , N_e ; \delta \apar , \delta N_e) =\intd d^2 x \,\left(\bstar \vert \nabla \delta \apar \vert^2 + \vert \nabla \delta \phi \vert^2 \right. \nonumber \\
& \left. -\lambda \vert \delta N_e\vert^2 +2 \F ' (\apar)  \delta N_e \delta \apar+ (\G '' (\apar) + N_e \F '' (\apar))  \vert \delta \apar \vert^2    \right).   \label{2fci}
 \end{align}
Considering $\delta(\F (\apar))=\F '(\apar) \delta \apar$ and using the boundary conditions (\ref{bcpertci}), the expression (\ref{2fci}) can be reformulated in the following way (see also Refs. \cite{Hol86b,Tro15}): 
 \begin{align}
& \delta^2 F(\apar , N_e ; \delta \apar , \delta N_e) =\intd  d^2 x \, \left( (\bstar - {\F '}^2 (\apar))\vert \nabla \delta \apar \vert^2 + \vert \nabla \delta \phi - \nabla \delta (\F (\apar))\vert^2 \right. \nonumber \\
& \left. + ( \F ' (\apar) \lapp \F ' (\apar) + \G '' (\apar) + N_e \F '' (\apar))\vert \delta \apar \vert^2 -\lambda \vert \delta N_e \vert^2 \right).  \label{2fci1}
\end{align}
We specialize now to the equilibria of interest and consider first the case $\F ' (\apar) \neq 0$. Making use of  the expressions (\ref{equi1ci})-(\ref{equi2ci}), as well as of the relations (\ref{Ff1})-(\ref{Gf1}), in Eq. (\ref{2fci1}), we obtain that the second variation, evaluated at the equilibrium of interest, can be rearranged to give
\begin{align}
& \delta^2 F(A_{eq} , K_1 \lapp A_{eq} ; \delta \apar , \delta N_e) = \nonumber \\
&\intd d^2 x \, \left(  (\bstar  - K_1^2 ( 1 - 2 \lambda \ee^{2 A_{eq}})^2)\vert \nabla \delta \apar \vert^2  + \vert \nabla \delta \phi -K_1(1-2 \lambda \ee^{2 A_{eq}})\nabla \delta \apar \vert^2  \right.  \label{2feqci}\\
& \left. +(\, K_1^2 - \bstar+4 \lambda K_1^2 \vert \nabla A_{eq} \vert^2 (2 \lambda \ee^{2 A_{eq}}-1)-4 \lambda^2 K_1^2 \ee^{4 A_{eq}} \, )2 \ee^{2 A_{eq}}\vert \delta \apar \vert^2 -\lambda \vert \delta N_e \vert^2  \right), \nonumber
\end{align}
where we also used the equilibrium relation $\lapp A_{eq}=-\exp(2 A_{eq})$.

The coefficients of $\vert\nabla \delta \apar \vert^2$, $\vert \delta \apar \vert^2$ and $\vert \delta N_e \vert^2$ in the integrand have indefinite sign. Identifying conditions for which they are positive will make the integrand, and in turn $\delta^2 F$, positive, thus providing stability conditions for the equilibria under consideration. We begin by noticing that $\lambda <0$ makes the coefficient of $\vert \delta N_e \vert^2$ positive. With regard to the coefficient of $\vert \nabla \delta \apar \vert^2 $, we observe that it is positive, on the domain, if 
\beq  \label{condci}
\bstar > \max_{(x,y)\in D_n} K_1^2(1 - 2 \lambda \ee^{2 A_{eq}(x,y)})^2.
\eeq
For $\lambda <0$, the maximum of the function on the right-hand side of Eq. (\ref{condci}) is attained at $x=\pi$ and $y=0$. Evaluating the function on the right-hand side of Eq. (\ref{condci}) at this point, yields the condition
\beq
\bstar > K_1^2 \left(1 - \frac{2 \lambda}{(a -\sqrt{a^2 -1})^2}\right)^2.
\eeq 
For $\lambda <0$ the coefficient of $\vert \delta \apar \vert^2$ is positive if $\bstar < K_1^2 (1-4 \lambda^2 \exp(4 A_{eq}))$. This condition, however, is in conflict with the condition (\ref{condci}). Again, we can resort to the Poincar\'e inequality (\ref{poinc}) which, if the condition (\ref{condci}) holds, when applied to the first term on the right-hand side of Eq. (\ref{2feqci}), provides the following bound:
\begin{align}
& \delta^2 F(A_{eq} , K_1 \lapp A_{eq} ; \delta \apar , \delta N_e)  \geq \intd d^2 x \, \left(\vert \nabla \delta \phi -K_1(1-2 \lambda \ee^{2 A_{eq}})\nabla \delta \apar \vert^2  \nonumber \right. \\
& \left. + (\kr(\bstar -K_1^2(1 - 2 \lambda \ee^{2 A_{eq}})^2) + ( \, K_1^2 - \bstar+4 \lambda K_1^2 \vert \nabla A_{eq} \vert^2 (2 \lambda \ee^{2 A_{eq}}-1) \right.   \label{boundci} \\
& \left. -4 \lambda^2 K_1^2 \ee^{4 A_{eq}} \, )2 \ee^{2 A_{eq}} ) \vert \delta \apar \vert^2 -\lambda \vert \delta N_e \vert^2  \right). \nonumber
\end{align}
The coefficient of $\vert \delta \apar \vert^2$ on the right-hand side of Eq. (\ref{boundci}) can be made positive considering that, for $\lambda <0$, the terms proportional to $\vert \nabla A_{eq} \vert^2$ are non-negative and noticing that
\begin{align}
& \kr(\bstar -K_1^2(1 - 2 \lambda \ee^{2 A_{eq}})^2) + ( \,K_1^2 - \bstar -4 \lambda^2  K_1^2 \ee^{4 A_{eq}} \, )2 \ee^{2 A_{eq}} \nonumber \\
& \geq \kr(\bstar -\max_{(x,y)\in D_n} \{ K_1^2(1 - 2 \lambda \ee^{2 A_{eq}(x,y)})^2 \})+ \min_{(x,y) \in D_n} \{(K_1^2 - \bstar -4 \lambda^2  K_1^2\ee^{4 A_{eq}} \, )2 \ee^{2 A_{eq}}\} \\
&=\kr\left(\bstar  - K_1^2 \left(1 -\frac{2\lambda}{(a -\sqrt{a^2 -1})^2}\right)^2\right) -2\frac{\bstar - K_1^2}{(a -\sqrt{a^2 -1})^2} -8 \frac{\lambda^2 K_1^2}{(a -\sqrt{a^2 -1})^6}.
\end{align}
We can therefore conclude that the second variation is positive, and consequently the equilibria (\ref{equi1ci})-(\ref{equi2ci}) are linearly stable, if the following three conditions are satisfied:
\begin{align}
&\bstar > K_1^2 \left(1 -\frac{2\lambda}{(a -\sqrt{a^2 -1})^2}\right)^2,  \label{cond1ci} \\
\nonumber \\  
& \lambda <0,   \label{cond2ci}\\
\nonumber  \\
&\left( \frac{1}{4 n^2}+\frac{\pi^2}{4 l^2}\right)\left(\bstar  - K_1^2 \left(1 -\frac{2\lambda}{(a -\sqrt{a^2 -1})^2}\right)^2\right) \nonumber \\
&> 2\frac{\bstar - K_1^2}{(a -\sqrt{a^2 -1})^2}+8 \frac{\lambda^2 K_1^2}{(a -\sqrt{a^2 -1})^6}. \label{cond3ci}
\end{align}
The condition (\ref{cond1ci}) can be seen as an upper limit, depending on $\bpe$, $\Theta_e$ and $a$, on the amplitude $K_1$ of the $\mathbf{E}\times\mathbf{B}$ flow. This condition also suppresses the firehose instability. The condition (\ref{cond2ci}), on the other hand, can be reformulated as 
\beq
\Theta_e < 1+\frac{2}{\bpe}
\eeq
and, analogously to Eq. (\ref{cond2hi}) of the hot-ion case, provides an upper bound on electron temperature anisotropy. One can note that, for $\Theta_e=1$, this condition is always satisfied. In this limit, the term $-\lambda \vert \delta N_e^2 \vert^2$ in Eq. (\ref{2feqci}) always provides a positive contribution to the second variation. This suggests that, for isotropic electron temperature, the electron pressure term associated with $\lambda$ has a stabilizing role, with respect to the reduced MHD case where $\lambda=0$. The condition (\ref{cond3ci}) can be fulfilled by sufficiently reducing the width of the islands letting the parameter $a$ approach $1$. Indeed, the left-hand side of Eq. (\ref{cond3ci}) can be made arbitrarily large letting $a \rightarrow 1^+$, in which limit $l \rightarrow 0^+$ and the term $\pi^2 / (4 l^2)$ goes to infinity. In the same limit, on the other hand, the denominators on the right-hand side tend to $1$, so that the right-hand side remains bounded. 

In the case $\F ' (\apar) =0$, the second variation, evaluated at the equilibria (\ref{eq1cib})-(\ref{eq2cib}), and using Eq. (\ref{Gcib}), reads
\begin{align}
& \delta^2 F(A_{eq} , \lapp \phi_{eq} ; \delta \apar , \delta N_e) = \nonumber \\
&\intd d^2 x \, \left(  \bstar  \vert \nabla \delta \apar \vert^2  + \vert \nabla \delta \phi \vert^2 - 2 \bstar \ee^{2 A_{eq}} \vert \delta \apar \vert^2 -\lambda \vert \delta N_e \vert^2 \right). \label{2feqcib}
\end{align}
The second variation (\ref{2feqcib}) actually corresponds to the second variation (\ref{2feqci}) in the limit $K_1=0$, i.e. with no $\mathbf{E}\times\mathbf{B}$ flow. However, as we pointed out in Sec. \ref{ssec:fvarci}, for $\F ' (\apar) =0$, the potential $\phi_{eq}$ can correspond to non-trivial flows. Nevertheless, stability conditions for this case can be directly obtained from Eqs. (\ref{cond1ci})-(\ref{cond3ci}) by setting $K_1=0$ and can be formulated as 
\begin{align}
& \frac{\bpe}{2+\bpe} < \Theta_e < 1 + \frac{2}{\bpe}, \label{cond1cib} \\
\nonumber  \\
&\left( \frac{1}{4 n^2}+\frac{\pi^2}{4 l^2}\right) > \frac{2 }{(a -\sqrt{a^2 -1})^2}. \label{cond2cib}
\end{align}
The condition (\ref{cond1cib}) comes from the requirements $\bstar >0$ and $\lambda <0$ and prevents  instabilities due to temperature anisotropy. The condition (\ref{cond2cib}), on the other hand, implies restrictions on $a$ and is amenable to the same considerations discussed for the condition (\ref{cond3bhi}) in the case with no perpendicular flow.

\section{Concluding remarks} \label{sec:concl}

In this work we studied the existence and the stability of stationary solutions, of a reduced fluid model, describing chains of magnetic vortices. The formation of chains of magnetic vortices, due to the reconnection of magnetic field lines, is a frequent phenomenon in laboratory and space plasmas. Observational evidence shows, in particular,  the existence of chains of magnetic vortices, for instance in the plasma of the solar wind. The presence of such structures can have a strong impact on the turbulent spectrum of magnetic and kinetic plasma energy. \\ 
We first reduced the general gyrofluid model, by acting on its Hamiltonian structure, to a 2D version without electron inertia effects.
Subsequently, we  considered the resulting model in the asymptotic limit in which the equilibrium ion temperature, referred to the plane perpendicular to the direction of a strong magnetic guide field, is much greater than the electron one, i.e. $\tau_{\pe_i} \gg 1$. In this limit we found equilibrium equations admitting solutions describing magnetic vortex chains supporting a class of non-trivial perpendicular flows, constant on the magnetic flux function contour lines,  and depending on an arbitrary function. We obtained that such magnetic vortex chains equilibria are linearly stable if three conditions are fulfilled. Two of these conditions  impose bounds on the electron temperature anisotropy, which, as expected, can be a source for instabilities. Depending on the range of values for $\bpe$, the temperature anisotropy has only a lower bound or is bounded from above and from below. Interestingly, the lower bound corresponds to the bound for firehose instability known for homogeneous equilibria according to linear wave stability analysis. Upper and lower bounds depend on the electron beta parameter. The third condition depends explicitly on the choice of the equilibrium flow. For a given flow and for fixed $\bpe$ and $\Theta_e$, it can be seen as a condition on the maximum island width and on the length of the chain. Shorter chains with thin islands favor stability. 

In the opposite, cold-ion case, with   $\tau_{\pe_i} \ll 1$, a slightly more intricate situation occurs, presenting two sub-cases. In one sub-case, the magnetic vortex chain supports an electrostatic potential $\phi$ linear with respect to the magnetic flux function. This restricts the equilibrium $\mathbf{E}\times\mathbf{B}$ velocity to be proportional to the local Alfv\'en velocity. The electron gyrocenter density $N_e$ and the parallel magnetic perturbations $\bpar$, on the other hand, are proportional to the current density associated with the vortex chain. In this sub-case, one stability condition suppresses the firehose instability but is stronger than the aforementioned condition, due to the presence of the equilibrium flow. A second condition sets an upper bound to temperature anisotropy and a third condition, again concerns also the size and the length of the chain. In the second sub-case, the fields $\phi$ and $N_e$ are no longer constrained to be constant on contour lines of the magnetic flux function and satisfy the relation $N_e = (-\phi +F_1) / \lambda$. In principle this can provide non-trivial flows.   Stability conditions bound temperature anisotropy from above and from below, with the lower bound again corresponding to the firehose stability condition. The third condition, on the other hand, turns out to correspond to the one found for the hot-ion case in the absence of flows. If a non-trivial solution for the flow can be found in this case, the characteristics of such solution appear not to be crucial for stability. 

Our analysis suggests that, in both hot and cold-ion regimes, several parameters of the system have to be controlled to attain the stability conditions. Such conditions appear to be rather compelling, and favor short chains with thin vortices and moderate anisotropy. The condition on the maximum vortex width is analogous to the condition for nonlinear stability of "cat's eyes" vortex chains derived in Ref.  \cite{Hol86b}.  We point out again, that our analysis is carried out over the domain enclosed by the separatrices and thus rules out external perturbations.  It is well known that magnetic island chains are actually unstable on larger domains including regions outside the separatrices \cite{Fin77,Pri79,Bon83b}. This seems to indicate that magnetic vortex chains might persist as coherent structures when perturbations coming from outside the chain are negligible.

Finally, it is appropriate to discuss some peculiarities and limitations of our approach based on the "cat's eyes"  solution. In our analysis, such solution was chosen for the magnetic equilibrium mostly because, as discussed in Sec. \ref{sec:intro}, it represents a classical (and one of the few) analytical two-dimensional equilibrium  solutions describing magnetic vortex chains. Imposing this equilibrium solution for $\apar$,  forced us to select  the free function $\G$ according to Eq. (\ref{ghi}) and (\ref{eq1cibis}) for the hot and  cold ion case, respectively. The existence of constants of motion characterized by arbitrary functions, as in Eq. (\ref{eq:c1}), is peculiar of 2D fluid systems. The consequent arbitrariness in the choice of the equilibria follows from the 2D symmetry. This in contrast with the 3D situation. As discussed in Ref. \cite{Yos03}, for instance, in 3D incompressible MHD, equilibria can be obtained from an analogous variational principle setting to zero the linear combination of three constants of motion  corresponding to the total energy, the magnetic helicity and the cross-helicity.  Such variational principle leads to the so-called Beltrami states, without arbitrariness due to the choice of free functions. It is also pointed out that, in 2D vortex dynamics, where, due to the presence of a symmetry, infinite constants of motion again appear, Beltrami states correspond to those obtained from a variational principle, in which the arbitrary function associated with the infinite number of invariants of motion, is chosen in such a way that the vorticity is a linear  function of the stream function. In this respect, therefore, the "cat's eyes" solutions is more of practical application (and of less fundamental physical significance as other analogous solutions, such as the Bernstein-Greene-Kruskal mode for the one-dimensional Vlasov-Poisson system). On the other hand, if one considers the "cat's eyes" solution (\ref{ceyes}), upon the substitution $\alpha=\sqrt{a^2 -1}$, and expanding about $\alpha=0$, one has
\beq   \label{ceyesexp}
A_{eq} (x,y)=-\log \cosh y -\alpha \frac{\cos x}{\cosh y}+\mathcal{O}(\alpha^2).
\eeq
The first term on the right-hand side of Eq. (\ref{ceyesexp}) corresponds to the classical Harris sheet equilibrium. The first order term can be seen as the external solution for the perturbation of the Harris sheet in the linear tearing stability problem on an infinite domain \cite{Whi86}. Therefore, the "cat's eyes" solution, in addition to "model" chains of magnetic islands, can be directly related to the linear reconnection problem as, in the limit $\alpha \rightarrow 0$, it corresponds to the superposition \cite{Cec05} of a Harris sheet with  the solution, of infinitesimal amplitude, of the linear problem, valid everywhere except at  the resonant surface. Such solution, in particular, yields $\Delta ' =0$, where $\Delta '$ indicates the standard tearing stability parameter \cite{Fur63}. This corresponds to a state of marginal stability. An interesting connection between the tearing mode solution and the variational principle leading to equilibria for MHD was developed in Ref. \cite{Yos12}. In this context the tearing mode solution is viewed as a singular equilibrium solution of MHD equations linearized about a Beltrami magnetic field satisfying the relation $\nabla\times \mathbf{B}=\mu\mathbf{B}$ with constant $\mu$. The equilibrium is obtained by extremizing a linear combination of the Hamiltonian and the Casimir invariant of the linearized system.  The corresponding Casimir invariant is referred to as "helical-flux Casimir". Violation of the conservation (due, for instance, to resistivity) of such Casimir invariant, leads to an instability that can allow to connect two Beltrami magnetic fields possessing the same magnetic helicity but belonging to two distinct categories, depending on whether the parameter $\mu$ belongs or not to the spectrum of the curl operator (defined on the appropriate functional space). This explains how a transition to a helical Beltrami field could occur. It might be of interest to investigate whether an analogous description for the tearing mode solution associated with the Harris sheet could be made and whether a possible relation exists with the "cat's eyes" solution for a finite value of $\alpha$. 

\section{Acknowledgments}

CG and ET acknowledge fruitful discussions with the members of the Fluid and Plasma Turbulence Team of the Laboratoire Lagrange.

\appendix

\section{Summary of the derivation of the model}  \label{sec:gyrokin}

 In this Section we describe the main steps in the derivation of the model from a system of gyrokinetic equations and discuss the underlying physical assumptions. Although in the previous stability analysis we made use of a 2D model, the following derivation will be slightly more general than required in order to show how, by properly choosing a scaling factor (indicated with $h(\epsilon)$ in Eq. (\ref{def2})), one can obtain the above adopted 2D model or an anisotropic 3D model. We also discuss the relation between the Hamiltonian structure of the model and that of a more general gyrofluid model presented in Ref. \cite{Tas19}.
 
 First, we recall the gyrokinetic system from which the reduced model (\ref{eq1red})-(\ref{ampperpred}) can be derived. This gyrokinetic system corresponds, in turn, to the gyrokinetic model derived in Ref. \cite{kunz2015} when equilibrium drifts are neglected and a bi-Maxwellian distribution is chosen as equilibrium distribution function. Such gyrokinetic system, in dimensional variables, reads 
\begin{align}
& \frac{\partial \ga}{\partial \tit}+\frac{c}{B_0}\left[ J_{0s} \wphi - \frac{\tv}{c} J_{0s} \wapar +2 \frac{\tmus B_0}{\qa}J_{1s}\frac{\wbpar}{B_0} , \ga \right]
 \nno \\
& +\tv \frac{\partial }{\partial \tz}\left( \ga  +\frac{\qa}{\Tpa} \tcalfa\left( J_{0s} \wphi - \frac{\tv}{c} J_{0s}\wapar +2 \frac{\tmus B_0}{\qa} J_{1s}\frac{\wbpar}{B_0}\right)\right)=0, \label{gyr}\\
&\sum_{s} \qa \int \dwa \, J_{0s} \ga = \sum_{s} \frac{\qa^2 }{\Tpea} \int \dwa \,  \tcalfa \left( 1 - J_{0s}^2 \right)  \wphi  \nno \\
&- \sum_s \qa \int \dwa \, 2 \frac{\tmus B_0}{\Tpea} \tcalfa J_{0s} J_{1s} \frac{\wbpar}{B_0},  \label{qndim}\\
&\sum_s \qa \int \dwa \, \tv J_{0s} \left( \ga -\frac{\qa}{\Tpa} \frac{\tv}{c} \tcalfa J_{0s} \wapar\right) \nno \\
&= -\frac{c}{4 \pi} \lapp \wapar + \sum_s \frac{\qa^2}{m_s}\int \dwa \,\tcalfa \left( 1 - \frac{1}{\thea}\frac{\tv^2}{\vtpa^2}\right)(1 - J_{0s}^2 ) \frac{\wapar}{c},  \label{ampdim}\\
&\sum_s \frac{\bepea}{n_0} \int \dwa \, 2 \frac{\tmus B_0}{\Tpea} J_{1s} \ga= - \sum_s \frac{\bepea}{n_0} \frac{\qa}{\Tpea} \int \dwa \,  2 \frac{\tmus B_0}{\Tpea} \tcalfa J_{0s} J_{1s}  \wphi  \nno \\
&-\left(2 + \sum_s \frac{\bepea}{n_0}\int \dwa \, \tcalfa \left( 2 \frac{\tmus B_0}{\Tpea} J_{1s}\right)^2 \right) \frac{\wbpar}{B_0}. \label{amppedim}
\end{align}
The index $s$ indicates the particle species ($s=e$ for electrons and $s=i$ for ions, when assuming a single ion species). Equations (\ref{qndim})-(\ref{amppedim}) express quasi-neutrality, parallel and perpendicular components of Amp\`ere's law, respectively. The gyrokinetic equation (\ref{gyr}) describes the evolution of the  function 
\beq  \label{pertdf}
\ga(\tx,\ty,\tz,\tv,\tmus ,\tit)=\dfa(\tx,\ty,\tz,\tv,\tmus,\tit)+\frac{\qa}{\Tpa}\frac{\tv}{c}\tcalfa (\tv , \tmus) J_{0s} \wapar (\tx,\ty,\tz,\tit).
\eeq
 In Eq. (\ref{pertdf}) $\dfa$ is the perturbation of the distribution function, whereas $\tcalfa$ is the bi-
Maxwellian equilibrium distribution function defined by
\beq  \label{bimax}
\tcalfa(\tv,\tmus)= \left(\frac{m_{s}}{{2 \pi}}\right)^{3/2} \frac{n_0}{\Tpa^{1/2} \Tpea}\mathrm{e}^{-\frac{m_{s} \tv^2}{2 \Tpa}-\frac{\tmus B_0}{ \Tpea }},
\eeq
The spatial variables $\tx,\ty,\tz$ are assumed to belong to the domain $\widetilde{\mathcal{T}}_n=\{(\tx,\ty,\tz) \in \mathbb{R}^3 \  | \ 0 \leq \tx \leq 2\pi  n \rhos, -\rhos L_y \leq \ty \leq \rhos L_y , -\lpar L_z \leq \tz \leq \lpar L_z\}$, with $n$ a non-negative integer and $L_y$ and $L_z$ two positive dimensionless constants. We indicated with $\lpar$, on the other hand, a length representing a characteristic scale of variation of the variable $\tilde{z}$. Periodic boundary conditions of $\ga$ assumed on this domain. Further independent variables are the time $\tit $,  the velocity coordinate parallel to the guide field $\tv$  and the magnetic moment (referred the unperturbed  magnetic guide field) $\tmus $ of the particle of species $s$.  We indicated with $m_s$ and $\qa$ the mass and the charge, respectively, of the particle of species $s$ and with $\dwa=(2 \pi B_0 / m_s) d \tmus d \tv$ the volume element in velocity space, integrated over the particle gyration angle. The gyroaverage operators $J_{0s}$ and $J_{1s}$ are defined as 
\begin{align}
&J_{0s} f (\tx,\ty,\tz )=\sum_{\tilde{\bk} \in \widetilde{\scrt_n}} J_0 (\aal) f_{\tilde{\bk}} \exp(i \tilde{\bk} \cdot \tilde{\bx}),   \label{j0}\\
&J_{1s} f (\tx,\ty,\tz)=\sum_{\tilde{\bk} \in \widetilde{\scrt_n}} \frac{J_1 (\aal)}{\aal} f_{\tilde{\bk}} \exp(i \tilde{\bk} \cdot \tilde{\bx}),  \label{j1}
\end{align}
where $J_0$ and $J_1$ are zero and first order Bessel functions of the first kind, $\aal=\tilde{k}_\perp \sqrt{2 \tmus B_0/ m_s} /\omega_{cs}$ is the perpendicular wave number multiplied times the gyroradius of the particle of species $s$ and $\tilde{\bx}$ a vector of coordinates $(\tx, \ty , \tz)$.  In Eqs. (\ref{j0})-(\ref{j1}), we indicated with $\widetilde{\scrt_n}$ the lattice $\widetilde{\scrt_n}=\{( l / ( n \rhos),  \pi m/ (\rhos L_y) ,  \pi p / (\lpar L_z) : (l,m,p) \in \mathbb{Z}^3 \}$.
 
 For the gyrofluid model we consider a plasma composed by electrons and a single ionized species of ions. \\
In order to perform a gyrofluid reduction of the gyrokinetic system (\ref{gyr})-(\ref{amppedim}), we introduce the following truncated Laguerre-Hermite expansion of the perturbation of the electron gyrocenter distribution function:
  \begin{align}
 & \dfe  (\tx,\ty,\tz,\tv,\tmue ,\tit)=\tcalfe (\tv , \tmue)\left(\frac{\widetilde{N}_e (\tx,\ty,\tz ,\tit)}{n_0}+\frac{\tv}{\vtpe}\frac{\widetilde{U}_e (\tx,\ty,\tz,\tit)}{\vtpe}  \right. \nno\\
 & \left. +\frac{1}{2}\left(\frac{\tv^2}{\vtpe^2}-1\right)\frac{\widetilde{T}_{ \parallel e} (\tx,\ty,\tz,\tit)}{\Tpe}+\left(\frac{\tmue B_0}{\Tpee}-1\right)\frac{\widetilde{T}_{ \perp e} (\tx,\ty,\tz,\tit)}{\Tpee}\right),  \label{decomp}
 \end{align}
 where $\vtpe=\sqrt{\Tpe / m_e}$, whereas $\widetilde{N}_e$, $\widetilde{U}_e$, $\widetilde{T}_{\parallel e}$ and $\widetilde{T}_{\perp e}$ represent the gyrofluid moments corresponding to fluctuations of the density, parallel velocity, parallel temperature and perpendicular temperature, respectively, all referred to the electron gyrocenters. These four fields, due to the orthogonality of Hermite and Laguerre polynomials, satisfy the relations
 \begin{align}
 & \frac{\widetilde{N}_e}{n_0}= \int \dwe \, \dfe, \qquad  \frac{\widetilde{U}_e}{\vtpe}= \frac{1}{n_0} \int \dwe \, \frac{\tv}{\vtpe} \dfe,  \nno \\
 & \frac{\widetilde{T}_{ \parallel e}}{\Tpe}=\frac{1}{n_0} \int \dwe \, \left( \frac{\tv^2}{\vtpe^2}-1 \right)\dfe, \qquad \frac{\widetilde{T}_{ \perp e}}{\Tpee}=\frac{1}{n_0} \int \dwe \, \left( \frac{\tmue B_0}{\Tpee}-1 \right)\dfe.   \label{defmom}
 \end{align} 
We introduce the parameters 
\begin{equation}
  \beta_{\pe_e}=8\pi \frac{ n_0 T_{0_{\perp e}}}{B_0^2}, \qquad   \tau_{\pe_i} = \frac{T_{0_{\perp  i}}}{T_{0_{\perp e}}},   \qquad   \Theta_e=\frac{T_{0_{\perp e}}}{T_{0_{ \parallel  e}}} \qquad \delta=\sqrt{\frac{m_e}{m_i}},  
\end{equation}
partially already defined in Eqs. (\ref{def}) and (\ref{param}), and the small parameter 
\beq
\epsilon=\frac{\rhos}{\lpar}.
\eeq
We assume the following ordering:
 \begin{align}
  &  \frac{1}{\omega_{ci}}\frac{\partial}{\partial \Tilde{t}}   \sim   \frac{\partial_{\tilde{z}} }{\partial_{\tilde{x}} }\sim \frac{\partial_{\tilde{z}} }{\partial_{\tilde{y}} } \sim  \frac{\widetilde{N}_e}{n_0}    \sim   \frac{\widetilde{U}_e}{c_{s \perp}}    \sim    \frac{\widetilde{T}_{\parallel e}}{\Tpe}  \sim     \frac{\widetilde{T}_{\perp e}}{\Tpee}  \nno \\
  & \sim    \frac{e \wphi}{T_{0_{\perp e}}}    \sim    \frac{\wapar}{B_0 \rhos}    \sim      \frac{\wbpar}{B_0}    \sim \epsilon \ll 1,   \label{ord1}\\
 &\frac{1}{n_0}\int \dwi \, J_{0i} \dfi \, \sim \, \frac{1}{n_0 c_{s \perp}}\int \dwi \, \tv J_{0i} \dfi \,  \sim \, \frac{1}{n_0} \int \dwi \, 2 \frac{\mu_{0i} B_0}{T_{0 \perp i}} J_{1i}\dfi \ll \epsilon,   \label{ord2}\\
 &\bpe \sim \tau_{\pe_i} \sim \Theta_e = \mathcal{O} (1),  \label{ord3}\\
 &\delta \ll 1,   \label{ord4}
 \end{align}
 and impose an isotropic equilibrium ion temperature, i.e.:
 \beq   \label{isotion}
 \Theta_i=1.
 \eeq
Concerning the physical motivations for the orderings (\ref{ord1})-(\ref{ord4}), we mention that the asymptotic relation concerning the first term in Eq. (\ref{ord1}) expresses the assumption of low-frequency fluctuations, whereas the assumption $\partial_{\tilde{z}} / \partial_{\tilde{x}} \sim \partial_{\tilde{z}} / \partial_{\tilde{y}} \sim \epsilon$ in Eq. (\ref{ord1}),  express anisotropy with respect to the direction of the magnetic guide field. Both assumptions belong to the customary gyrokinetic and gyrofluid orderings. The remaining relations in Eq. (\ref{ord1}), on the other hand, imply small fluctuations of the electromagnetic fields and of the involved gyrofluid electron moments, with respect to characteristic values taken as reference. Eq. (\ref{ord2}) allows to neglect ion gyrocenter  fluctuations in Eqs. (\ref{qndim})-(\ref{amppedim}), effectively decoupling the ion gyrocenter  dynamics.   Eq. (\ref{ord3}) allows to keep finite values of perpendicular electron and ion equilibrium temperatures as well as electron equilibrium temperature anisotropy. Finally, Eq. (\ref{ord4}) amounts to neglect electron inertia effects and, in particular, phenomena occurring at the scale of the electron skin depth and of the electron Larmor radius. Magnetic chains on scales much larger than such scales are indeed those observed in the solar wind \cite{perrone2017,Jov18}. The condition (\ref{isotion}), on the other hand, is imposed at this stage mainly for simplicity, the focus of the analysis being mostly on the electron species. 
We introduce the following set of normalized variables
\begin{equation} \label{def2}
\begin{split}
  &  
  x=\frac{\Tilde{x}}{\rhos}, \qquad    y=\frac{\Tilde{y}}{\rhos},  \qquad z= h(\epsilon) \frac{\Tilde{z}}{\lpar}, \qquad  t=\epsilon \, \omega_{ci}  \Tilde{t} ,  \\
  & N_e=\frac{1}{\epsilon}\frac{\widetilde{N}_e}{n_0 },   \qquad \apar= \frac{1}{\epsilon} \frac{\wapar}{B_0 \rhos},    \\
   & \phi=\frac{1}{\epsilon} \frac{e \wphi}{T_{0_{\perp e}}},   \qquad   U_e= \frac{1}{\epsilon} \frac{\widetilde{U}_e}{c_{s \perp}},   \qquad    \bpar=\frac{1}{\epsilon} \frac{\wbpar}{B_0},\\
   & T_{\parallel e}=\frac{1}{\epsilon} \frac{\widetilde{T}_{\parallel e}}{\Tpe}, \qquad T_{\perp e}=\frac{1}{\epsilon} \frac{\widetilde{T}_{\perp e}}{\Tpee}
   \end{split}  
\end{equation}
 Inserting the parameter $\epsilon$ in the normalization enables us to obtain variables according to which all  terms in the resulting equations will be of order unity, with the exception of the terms including partial derivatives with respect to $z$, for which the function $h(\epsilon)$ will determine the order. The function $h(\epsilon)$  is taken to be an arbitrary function, at most of order $1$ as $\epsilon$ tends to $0$. By taking $h(\epsilon) \rightarrow 0$ as $\epsilon \rightarrow 0$, the derivatives with respect to $z$ will turn out to be negligible and consequently we will obtain the 2D model. On the other hand, by taking $h(\epsilon) \sim 1$ as $\epsilon \rightarrow 0$, the system will remain 3D.  \\ 
We remark that, on the basis of the relations (\ref{ord1}), the expression for the normalized magnetic field $\mathbf{B}=\tilde{\mathbf{B}}/B_0$ (with $\tilde{\mathbf{B}}$ indicating the dimensional magnetic field) reads
 \begin{equation} \label{magfield}
    \mathbf{B}(x,y,z,t) =  \hz + \epsilon \bpar(x,y,z,t)\hz + \epsilon \nabla \apar (x,y,z,t)\times \hz + \mathcal{O}(\epsilon^2).
\end{equation}
 The first term on the right-hand side of Eq. (\ref{magfield}) corresponds to the constant and uniform guide field, which is of order unity. The next two terms, both of order $\epsilon$, correspond to the fluctuations of the parallel and perpendicular magnetic field. The condition $\nabla\cdot\mathbf{B}=0$ implies that the right-hand side of Eq. (\ref{magfield}) has to be divergence-free. The contribution $\partial_{z} \bpar$ to $\nabla\cdot \mathbf{B}$ is in general not zero but is, however, of order $\epsilon^2 h(\epsilon)$ and thus at most of order $\epsilon^2$. The exact expression for the terms of order $\epsilon^2$ or higher, in the magnetic field (including those that would contribute to give $\nabla\cdot\mathbf{B}=0$ compensating $\partial_{z} {\bpar}$) is, however, not required as, for the present theory, such terms always yield contributions that are of higher order with respect to those retained in the adopted gyrokinetic and gyrofluid model equations.

In order to proceed with the derivation, we insert the decomposition (\ref{decomp})   into the parent system (\ref{gyr})-(\ref{amppedim}). Making use of the orthogonality of the Hermite polynomials and of the relations (\ref{defmom}) one can obtain, from Eq. (\ref{gyr}) evolution equations for the gyrofluid moments.  Transforming to the dimensionless variables (\ref{def2}) and applying the limit (\ref{ord4}), one obtains the following evolution equations for the fields $N_e$ and $\apar$:
\begin{align}
   & \frac{\partial N_e}{\partial t}+[\phi, N_e] - [\bpar, N_e + T_{\perp e}] - [\apar , U_e] + h(\epsilon)\frac{\partial U_e}{\partial z}=0,
    \label{eq:mod1} \\
   & \frac{\partial \apar}{\partial t} + [\phi - \bpar, \apar  ] + \frac{1}{\Theta_e}[\apar, N_e + T_{\parallel e}] +h(\epsilon) \frac{\partial}{\partial z}\left( \phi - \bpar - \frac{N_e}{\Theta_e}- \frac{T_{\parallel e}}{\Theta_e}\right) =0.
    \label{eq:mod2}
\end{align}
On the other hand, by the same procedure, one obtains, from Eqs. (\ref{qndim})-(\ref{amppedim}), the following leading order relations:
\begin{align}
   &  N_e + (1 - \Gamma_{0i} + \Gamma_{1i}) \bpar + (1- \Gamma_{0i} ) \frac{\phi}{\tau_{\pe _i}} =  0, \label{qntemp}\\
   &  U_e  =  b_{\star} \Delta_{\perp} \apar,  \label{amppartemp}\\
   &  \bpe(N_e + T_{\perp e})=\bpe(1-(\Gamma_{0i} - \Gamma_{1i}))\phi-2\bpar -2\bpe(1+\tau_{\perp_i}(\Gamma_{0i}-\Gamma_{1i}))\bpar.  \label{ampperptemp}
\end{align}
We note that in Eqs. (\ref{qntemp}) and (\ref{ampperptemp}) the gyroaverage operators $\Gamma_{0i}$ and $\Gamma_{1i}$, defined in Eqs. (\ref{gamma0})-(\ref{gamma1}), descend from the integrals in Eqs. (\ref{qndim}) and (\ref{amppedim}) involving the Bessel functions $J_0$ and $J_1$.

The system composed by Eqs. (\ref{eq:mod1})- (\ref{ampperptemp})  is evidently not closed, as no evolution equations for the temperature fluctuations $\tpe$ and $\tpee$ have been provided. Instead of deriving such equations, we impose closure relations, postponing an analysis of temperature and kinetic effects on stability, to future work. In particular, we close the system by imposing   
\beq  \label{closure}
\tpe=0, \qquad \tpee=-\bpar.
\eeq
The relations (\ref{closure}) correspond to imposing isothermal closures for both the parallel and perpendicular electron (instead of electron gyrocenter) temperatures. Indeed, when electron FLR effects are neglected, as in this case, the perpendicular and parallel temperature electron fluctuations $t_{\perp e}$ and $t_{\parallel e}$ are related to those of electron gyrocenters $\tpee$ and $\tpe$, by $t_{\perp e}=\tpee+\bpar$ and $t_{\parallel e}=\tpe$, as explained in Ref. \cite{Bri92} (although the latter Reference treats the isotropic case $\Theta_e=1$, such difference is irrelevant for the relations of use in this case).\\

The resulting system reads
\begin{align}
    &\frac{\partial N_e}{\partial t}+[\phi - \bpar, N_e] - [\apar , U_e] +h(\epsilon) \frac{\partial U_e}{\partial z}=0,
    \label{cont1} \\
&    \frac{\partial \apar}{\partial t} + [\phi - \bpar, \apar  ] + \frac{1}{\Theta_e}[\apar, N_e ] +h(\epsilon) \frac{\partial}{\partial z}\left( \phi - \bpar - \frac{N_e}{\Theta_e}\right) =0,   \label{mom1} \\
&  N_e+ (1 - \Gamma_{0i} + \Gamma_{1i})\bpar + (1- \Gamma_{0i} ) \frac{\phi}{\tau_{\pe _i}} =  0, \label{qn1}\\
   &  U_e  =  b_{\star} \Delta_{\perp} \apar,  \label{amppar1}\\
  & \bpar= - \frac{\bpe}{2}(N_e-(1 - \Gamma_{0i} + \Gamma_{1i})\phi + (1+ 2 \tau_{\perp_i}(\Gamma_{0i} - \Gamma_{1i}))\bpar).  \label{ampperp1}
\end{align}
As previously stated, choosing $h$ such that $h(\epsilon) \rightarrow 0$ as $\epsilon \rightarrow 0$, leads to neglecting the terms with the derivatives with respect to $z$, in such manner that the model is reduced to the 2D system. In this limit, Eqs. (\ref{cont1})-(\ref{ampperp1}) correspond indeed to Eqs. (\ref{eq1red})-(\ref{ampperpred}), which are the model equations adopted in our analysis. We also note that, for such choice of $h$, the spatial domain of the fields $N_e$, $\apar$, $\phi$, $U_e$ and $\bpar$ in Eqs. (\ref{cont1})-(\ref{ampperp1})  is given by $\mathcal{D}_{n0}=\{(x,y,z) \in \mathbb{R}^3 \  | \ 0 \leq x \leq 2\pi n , -L_y \leq y \leq L_y , z=0 \}$, obtained by taking the $\epsilon \rightarrow 0$ limit of the domain $ \{(x,y,z) \in \mathbb{R}^3 \  | \ 0 \leq x \leq 2\pi n , -L_y  \leq y \leq L_y , -h(\epsilon) L_z \leq z \leq h(\epsilon) L_z \}$, corresponding to the domain $\widetilde{\mathcal{T}}_n$ written in terms of the dimensionless variables (\ref{def2}). The domain $\mathcal{D}_{n0}$ is isomorphic to $\dn$. Therefore,  the 2D model described in Sec. \ref{sec:model} can be obtained as a leading order expansion, as $\epsilon\rightarrow0$, of the model (\ref{cont1})-(\ref{ampperp1}) when the variable $z$ is sufficiently "contracted" by choosing $h(\epsilon) \ll 1$.  On the other hand, if one chooses $h$ such that $h(\epsilon ) \sim 1 $ as $\epsilon \rightarrow 0$, the terms involving $z$-derivatives in Eqs. (\ref{cont1})-(\ref{mom1}) remain of order unity, as the other terms in the equations. In this case one obtains a 3D model with fields defined over the domain $ \{(x,y,z) \in \mathbb{R}^3 \  | \ 0 \leq x \leq 2\pi n , -L_y  \leq y \leq L_y , - L_z \leq z \leq  L_z \}$.\\ 
We remark that, up to the normalization, one could alternatively obtain the system (\ref{eq1red})-(\ref{ampperpred}) and its Hamiltonian structure, from the system \cite{Tas19}
 \begin{equation}
    \frac{\partial N_e}{\partial t}+[\phi, N_e] - [B_{\parallel}, N_e] - [A_\parallel , U_e] + \frac{\partial U_e}{\partial z}=0,
    \label{modgen1}
\end{equation}
\begin{equation}
    \frac{\partial}{\partial t}( A_\parallel - \delta^2 U_e ) + [\phi - B_{\parallel}, A_\parallel - \delta^2 U_e ] + \frac{1}{\Theta_e}[A_\parallel,N_e] + \frac{\partial}{\partial z}\left( \phi - B_\parallel - \frac{N_e}{\Theta_e}\right) =0,
    \label{modgen2}
\end{equation}
complemented by the static relations 
\begin{align}
   &  N_e + (1 - \Gamma_{0i} + \Gamma_{1i})B_\pa + (1- \Gamma_{0i} - \tau_{\pe_i}\delta^2 \Delta_{\pe}) \frac{\phi}{\tau_{\pe _i}} =  0, \label{26}\\
   &  U_e  =  b_{\star} \Delta_{\perp} A_{\parallel},  \label{amppar}\\
   &  B_{\parallel}= - \frac{\bpe}{2}(N_e -(1 - \Gamma_{0i} + \Gamma_{1i})\phi + (1+ 2 \tau_{\perp_i}(\Gamma_{0i} - \Gamma_{1i}))B_{\parallel}).  \label{ampperp}
\end{align}
Written with the appropriate normalization, this gyrofluid model permits to extend some reduced fluid models present in the literature.
For instance, it extends the model derived in Ref. \cite{Pas18} by adding equilibrium electron temperature anisotropy. 
 In the limit $\Theta_e=1$, $\tau_{\perp_i }\ll 1$ and for negligible parallel magnetic perturbations, it reduces to the  two-field model considered in a number of works on collisionless magnetic reconnection such as those of Refs. \cite{Sch94} and \cite{Caf98}. The model can also be seen as an extension of the model for inertial kinetic Alfv\'en turbulence described in Ref. \cite{Che17}, accounting also for ion finite Larmor radius effects, parallel electron pressure and equilibrium electron temperature anisotropy.  
 
The system (\ref{modgen1})-(\ref{ampperp})  was shown to be Hamiltonian in Ref. \cite{Tas19}. Its Hamiltonian structure consists of the Hamiltonian functional
\begin{equation}
    H(N_e, A_e)= \frac{1}{2} \int_{\mathcal{T}_n}  d^3 x \, \left( \frac{N_e^2}{\Theta_e} -\bstar A_e \lapp \lba A_e  -N_e \lbphi N_e +N_e \lbb N_e  \right),
    \label{eq:ham}
\end{equation}
and of the Poisson bracket 
\begin{equation}
\begin{split}
     & \{ F , G \}= \int_{\mathcal{T}_n}  d^3 x \, \Bigg( N_e \left(  [F_{N_e}, G_{N_e}]  + \frac{\delta^2}{\Theta_e}[F_{A_e} , G_{A_e}]\right)  \\
     & + A_e([F_{A_e} , G_{N_e}] + [F_{N_e} , G_{A_e}]) + F_{N_e} \frac{\partial G_{A_e}}{\partial z} + F_{A_e} \frac{\partial G_{N_e}}{\partial z} \Bigg).
    \label{eq:pb}
\end{split}
\end{equation}
In Eqs. (\ref{eq:ham}) and (\ref{eq:pb}) $A_e=A_\pa - \delta^2 U_e$ whereas $\lba$, $\lbphi$ and $\lbb$ are linear operators that permit to express $\apar$, $\phi$ and $\bpar$ in terms of the dynamical variables $N_e$ and $A_e$ by means of Eqs. (\ref{26})-(\ref{ampperp}). More precisely, the relations $\apar = \lba A_e$, $\phi = \lbphi N_e$, $\bpar= \lbb N_e$ hold. In Fourier space, such operators amount to Fourier multipliers. Analogously to the operators $\lphi$ and $\lb$ introduced in Sec. \ref{ssec:hamstruct}, also  $\lba$, $\lbphi$ and $\lbb$ are symmetric with respect to an appropriate inner product, which permits to show the conservation of $H$. 

The Hamiltonian (\ref{hamred}) and the Poisson bracket (\ref{pbred}) of the model used for our stability analysis can be obtained from the Hamiltonian (\ref{eq:ham}) and the Poisson bracket (\ref{eq:pb}) by setting $\delta=0$ and by restricting the algebra of observables to functionals of the dynamical variables $N_e=N_e(x,y,t)$ and $\apar=\apar(x,y,t)$ with $(x,y)\in \dn$. We remark that these operations do not spoil the Hamiltonian structure as, in particular, the Poisson bracket (\ref{eq:pb}) satisfies the Jacobi identity for any value of $\delta$, and in particular for $\delta=0$. This permits to obtain the target model while guaranteeing that the Hamiltonian character of the parent model does not get violated  in the reduction process. On the other hand, we point out that, from the point of view of the ordering, the model (\ref{modgen1})-(\ref{ampperp}) assumes $\bpe \sim \delta$, as in its version with isotropic electron temperature described in Ref. \cite{Pas18}. This indeed allows for retaining electron inertia contributions while neglecting most of electron FLR corrections. When, however, electron inertia is neglected as well, as in the target model, this hypothesis should be relaxed, by letting $\bpe =\mathcal{O}(1)$.

 \section{The stability algorithm}  \label{sec:stab}
 
 In this Section we briefly summarize the steps required for determining linear stability conditions   according to the Energy-Casimir method.
 
 We consider a dynamical system

 \beq  \label{dynsys}
 \frac{\partial \chi_i}{\partial t}=X_i (\chi_1, \cdots, \chi_N), \quad i=1, \cdots , N,
 \eeq
 evolving $N$ fields $\chi_1, \cdots , \chi_N$ all of which depend on time and on space variables $x_1, \cdots , x_m$ belonging to some domain $U \subset \mathbb{R}^m$, with $m$ and $N$ positive integers.
 
 We suppose the system admits a family of $s$ constants of motion $\mathcal{C}_1, \cdots , \mathcal{C}_s$, i.e. functionals $\mathcal{C}_1 (\chi_1, \cdots , \chi_N), \cdots ,\mathcal{C}_s (\chi_1, \cdots, \chi_N)$
 such that $d \mathcal{C}_i /dt=0$, for $i=1, \cdots , s$.
 The functional $F=\sum_{i=1}^s \mathcal{C}_i$ is then a constant of motion as well. For noncanonical Hamiltonian systems, a natural choice for $F$ is given by $F=H+\sum_{i=1}^{s-1}C_i$, where $H$ is the Hamiltonian of the system and $C_1, \cdots , C_{s-1}$ are Casimir invariants. This is why the method is referred to as Energy-Casimir method. 
 
 Solutions of the equation
 \beq
 \delta F (\chi_1, \cdots, \chi_N ; \delta \chi_1, \cdots , \delta \chi_N)=0,
 \eeq
 where $\delta F$ is the first variation of $F$, correspond to equilibria of the system (\ref{dynsys}). Such equilibrium points, denoted as $(\chi_{e1}, \cdots , \chi_{eN})$,  can then be related to constants of motion by requiring that    $(\chi_{e1}, \cdots , \chi_{eN})$ be a point where $\delta F$ vanishes. In this way, classes of equilibria (although in general not all the equilibria of the system) can be associated with different choices of constants of motion. 
 
 An equilibrium $(\chi_{e1}, \cdots , \chi_{eN})$ solution of $\delta F(\chi_1, \cdots, \chi_N ; \delta \chi_1, \cdots , \delta \chi_N)=0$ is formally stable (which implies linearly stable) if the second variation of $F$, evaluated at such equilibrium, i.e.
 \beq  \label{secvar}
 \delta^2 F (\chi_{e1}, \cdots, \chi_{eN} ; \delta \chi_1, \cdots , \delta \chi_N)
 \eeq
 has a definite sign. If this is the case, in fact, the expression (\ref{secvar}) (or its opposite) can be taken as a conserved norm for the system (\ref{dynsys}) linearized about the equilibrium $(\chi_{e1}, \cdots , \chi_{eN})$.
 
  \section*{References}

\bibliographystyle{iopart-num}
\bibliography{bibliorevised}

\end{document}